\documentclass[fleqn,usenatbib]{mnras}
\usepackage[english]{babel}
\usepackage[T1]{fontenc}
\usepackage{ae,aecompl}
\usepackage{fancyhdr}
\usepackage{bm}
\usepackage{amsfonts}
\usepackage{amsmath}
\usepackage{amssymb}
\usepackage{multicol}
\usepackage{layout}
\usepackage{graphicx}
\usepackage{multirow}
\usepackage{color}
\usepackage{multirow}
\usepackage{times}
\usepackage{natbib}
\def\be{\begin{equation}}
	\def\ee{\end{equation}}

\usepackage{lineno}

\usepackage[utf8]{inputenc}
\usepackage{lmodern}

\definecolor{valecol}{rgb}{0,0.5, 1.}

\newif\ifAMStwofonts
\AMStwofontstrue

\graphicspath{{./fig/}}
\title[ $\Lambda$CDM model against cosmography]{$\Lambda$CDM model against cosmography: A possible deviation after DESI 2024}

\author[Pourojaghi, Malekjani and Davari]{
	Saeed Pourojaghi$^{1}$, Mohammad Malekjani \thanks{malekjani@basu.ac.ir}$^{1}$ and Zahra Davari$^{2}$ \\ 
	$^1$ Department of Physics, Bu-Ali Sina University, Hamedan 65178, Iran\\
	$^2$ School of Physics, Korea Institute for Advanced Study (KIAS), 85 Hoegiro, Dongdaemun-gu, Seoul, 02455, Korea}
\date{Accepted ?, Received ?; in original form \today}
\pagerange{\pageref{firstpage}--\pageref{lastpage}} \pubyear{2024}

\begin{document}
\label{firstpage}
\maketitle		
\begin{abstract}
In this study, we present an analysis of the standard flat-$\Lambda$CDM model using a cosmographic approach, incorporating recent DESI BAO observations and Supernovae Type Ia catalogues (SNIa), including the DES-SN5YR and Pantheon+ compilations. We find full consistency between the standard model and the cosmographic approach when considering DESI BAO and SNIa catalogues independently.
When combining DESI BAO with SNIa data, we examine the impact of the Planck prior on the sound horizon at the drag epoch, $r_d$, and the Cepheid prior on the absolute magnitude, $M$. Applying the Planck prior on $r_d$ alone yields an $H_0$ value consistent with the 
Planck measurement, while applying the Cepheid prior on $M$ alone results in an $H_0$ value consistent with the SH0ES measurement.
Without any priors, the $H_0$ value obtained has a large error margin, reconciling the Planck and SH0ES measurements. In all cases where individual priors are applied, we observe no significant tension between the flat-$\Lambda$CDM model and the cosmographic approach. However, when both Planck and Cepheid priors are applied simultaneously, significant tensions arise between the model and cosmography. This tension is even more pronounced when excluding LRG1 and LRG2 from the DESI measurements. These results indicate that the standard model cannot simultaneously reconcile high-redshift Planck CMB observations and local Cepheid measurements. This discrepancy supports the possibility of new physics beyond the standard model or, alternatively, the presence of unrecognized systematic errors in the observational data.\\
\newline
\textbf{Keywords}: Cosmology: dark energy - Cosmology: cosmological parameters - Cosmology: theory.
\end{abstract}

\maketitle
\section{Introduction}
Observational evidence from distant Type Ia Supernovae (SNIa) \citep{Riess:1998cb,Perlmutter:1998np} indicates that our Universe has entered an accelerated phase of expansion. This cosmic acceleration suggests the presence of dark energy (DE) with negative pressure within the framework of General Relativity (GR). For comprehensive reviews on DE, see \citep{Frieman:2008sn,Weinberg_2013}. The initial groundbreaking results from SNIa observations have been consistently confirmed by increasingly accurate observations of SNIa at higher redshifts \citep{SupernovaSearchTeam:2004lze,ESSENCE:2007acn,Conley_2010,Suzuki_2012}. Additionally, high-precision observations of the anisotropies in the cosmic microwave background (CMB) radiation \citep{Komatsu2009,Jarosik:2010iu,Ade:2015rim,Planck:2018vyg}, weak gravitational lensing \citep{Benjamin:2007ys,Amendola:2007rr,Fu:2007qq}, baryon acoustic oscillations (BAO), and the large-scale structure of the Universe \citep{Tegmark:2003ud,Cole:2005sx,Eisenstein:2005su,Percival2010,Blake:2011en,Reid:2012sw} further substantiate the current positive acceleration phase of the Universe’s expansion.
A spatially flat universe with a present energy budget consisting of approximately $5\%$ baryonic matter, $25\%$ cold dark matter (CDM), and $70\%$ DE in the form of Einstein's cosmological constant $\Lambda$, with a negligible contribution from radiation, is known as the standard cosmological model or flat-$\Lambda$CDM model \citep{Peebles:2002gy}. However, the standard $\Lambda$CDM model faces both theoretical and observational challenges. Theoretically, it grapples with issues such as the fine-tuning and cosmic coincidence problems \citep{Weinberg:1988cp,Sahni:1999gb,Carroll:2000fy,Padmanabhan:2002ji}. Observationally, it encounters puzzling issues like the Hubble tension and the $S_8$ tension \citep{Perivolaropoulos:2021jda}. The persistence of these issues, supported by increasingly accurate cosmological observations, has motivated cosmologists to explore alternative models and new physics beyond the standard $\Lambda$CDM cosmology.
The first alternative suggests dynamical DE models with a time-varying equation of state (EoS) parameter, differing from the constant EoS parameter of the $\Lambda$CDM model ($w_{\Lambda}=-1$) within the context of GR \citep[For a review, see][]{Copeland:2006wr}. The second alternative proposes a fundamental revision of GR, introducing a new framework known as modified gravity theories (For a review, see \citep{Tsujikawa:2010zza}). It is obvious that the dynamics of the accelerated expanding universe depends on the specific DE model via the Friedmann equation. Alternatively, the cosmographic approach allows us to describe the expansion of the Universe independently of any specific DE model. In this approach, the Hubble parameter is expressed as an expansion in terms of redshift through mathematical approximations, detailed in Sect. \ref{sect:cosmography}. The coefficients of this expansion are directly related to the time derivatives of the scale factor, known as cosmographic parameters. Numerous studies have analyzed various cosmological models, including the standard $\Lambda$CDM model, using the cosmographic approach \citep{Sahni:2002fz, Alam:2003sc,Cattoen:2007sk, Capozziello:2009ka, Capozziello:2011tj, Capozziello:2018jya, Benetti:2019gmo, Escamilla-Rivera:2019aol, Lusso:2019akb, Rezaei:2020lfy, Bargiacchi:2021fow, Capozziello:2021xjw,Bargiacchi:2023rfd, Bamba:2012cp, Mandal:2020buf, Mishra:2024oln, Aviles:2013zz, Dunsby:2015ers}.
Utilizing observational datasets, cosmologists place constraints on cosmographic parameters, particularly the first two: the deceleration parameter ($q_0$) and the jerk parameter ($j_0$), to examine cosmological models. Cosmography is a robust method for testing potential deviations from the standard $\Lambda$CDM model at low redshifts. Cosmography has been employed to investigate the $H_0$ and $\sigma_8$ tensions observed in $\Lambda$CDM cosmology \citep{DAgostino:2023cgx}. They have demonstrated that cosmography can mitigate these tensions more effectively than the estimates derived from $\Lambda$CDM cosmology. The cosmographic approach has also been utilized to investigate potential violations of the Cosmic Distance Duality Relation (CDDR). Within this framework, \cite{Jesus:2024nrl} did not observe significant deviations from the CDDR. Their results remain consistent with the predictions of $\Lambda$CDM cosmology. 
Recent study \citep{Lusso:2019akb}, has used Hubble diagrams of Quasars (QSOs) Gamma Ray Bursts (GRBs) and SNIa to constrain cosmographic parameters, revealing significant tension between the standard $\Lambda$CDM model and the cosmographic approach. Their approach is based on the logarithmic expansion of the luminosity distance in terms of $z$. Additionally, they used another expansion, which is a Taylor series around $y = \frac{z}{1+z}$ \citep[see also][]{Risaliti:2018reu}. Furthermore, \citep{Bargiacchi:2021fow} applied orthogonalized logarithmic polynomials of the luminosity distance and found a significant tension ($>4\sigma$) between the $\Lambda$CDM model and the cosmographic approach using Hubble diagrams of SNIa and QSOs \citep[see also][]{Bargiacchi:2023rfd}.
While these reported tensions in cosmography are intriguing, it is essential to consider the error propagation of cosmography at higher redshifts. In this concern, \cite{Yang:2019vgk} demonstrated that the logarithmic polynomial expansion generally fails to recover the flat $\Lambda$CDM beyond $z \sim 2$. Additionally, they argued that the approximation in the $y$-expansion is not accurate at $z = 2$ with only five terms in the expansion \citep[see also][]{Cattoen:2007sk, OColgain:2021pyh}. Indeed, deviation between the cosmological model and cosmography may arise from truncation error of the mathematical approximations used in cosmography. Mock analyses have shown that the tension reported in \citep{Lusso:2019akb} is not physical and results from the inadequacy of cosmography at higher redshifts than the observed redshifts of SNIa where GRBs and QSOs have been observed. More recently, \cite{Pourojaghi:2022zrh} have demonstrated there is no tension between the $\Lambda$CDM model and cosmography, even at the high redshifts of GRB and QSO observations. Their mock analysis emphasized the importance of validating the cosmographic approach before application. Specifically, using the Padé approximation or Taylor expansion up to the 5th order of $y$-redshift, they showed the $\Lambda$CDM cosmographic parameters are consistent with those of cosmography even beyond the redshifts of SNIa \citep{Pourojaghi:2022zrh}.\\
While our focus is not on GRBs and QSOs as high-redshift observations, we emphasize that the Padé-cosmography method, a cosmography based on rational Padé-polynomials, has been well-examined and works effectively at higher redshifts beyond those of SNIa observations. We can safely use Padé-cosmography at redshifts smaller than $z\simeq 2.5$, where we aim to utilize the recent updated samples of SNIa data, including Pantheon+ \citep{Scolnic:2021amr} and DES-SN5YR \citep{DES:2024tys}, along with high-precision BAO measurements from the Dark Energy Survey Instruments team (DESI) \citep{DESI:2024mwx}.
Recently, \cite{DES:2024ywx} have measured the Hubble constant $H_0$ using DES-SN5YR and DESI BAO datasets, where the absolute magnitude of SNIa is calibrated by DESI BAO measurements. Interestingly, their preferred value of
$H_0$ agrees with the best-fitting Planck $H_0$ value in flat-$\Lambda$CDM model. This result is obtained by using the Planck prior on the sound horizon, $r_d$, at the time of photon-baryons decoupling after recombination.
In this paper, we explore potential deviations between the $\Lambda$CDM model and the cosmographic approach using DESI BAO measurements in combination with SNIa samples, including Pantheon+ and DES-SN5YR. We employ cosmography based on rational Padé polynomials which is effective at the redshifts of observed SNIa and DESI BAO measurements. For a similar study in the context of Taylor series see \citep{Luongo:2024fww}. Using the Markov Chain Monte Carlo (MCMC) algorithm, we will find the best fit of cosmographic parameters within $1-3\sigma$ uncertainties by minimizing the $\chi^{2}$ function.
Importantly, we investigate the impact of Cepheid priors (Calibrating the SNIa's absolute magnitude, $M$, using Cepheids in Pantheon+ sample \citep{Scolnic:2021amr}) and Planck prior for the sound horizon at the time of photon-baryons drag, $r_d$ on our analysis. 
Additionally, we perform the same analysis within the $\Lambda$CDM model to put the same observational constraints on cosmographic parameters. Interestingly, we will show a significant tension appears between the cosmographic parameters $q_0$ and $j_0$ of the $\Lambda$CDM model and those of cosmography when we use the Cepheid prior for $M$ and the Planck prior for $r_d$ at the same time.\\
The structure of this paper is organized as follows: In Sect. \ref{sect:cosmography}, we briefly review the Padé-cosmography and $\Lambda$CDM parameters in the cosmography context. In Sect. \ref{sec:data}, the observational data used in our analysis are introduced. In Sect. \ref{sec:data-analysis}, the numerical results are presented. Finally, we conclude our work in Sect. \ref{conlusion}.
\section{The Cosmographic Approach} \label{sect:cosmography}
Cosmography is a model-independent method that explains the history of the universe without assuming any specific cosmological model. This approach is based solely on the assumptions of homogeneity and isotropy of the universe. Within the framework of the Friedman-Lemaître-Robertson-Walker (FLRW) metric, cosmographic parameters can be expressed through the derivatives of the scale factor, $a$, with respect to cosmic time as follows \citep{Visser:2003vq}:
\begin{equation}\label{c1}
	\begin{aligned}
		H(t) &= \frac{1}{a} \frac{da}{dt}, \\
		q(t) &= -\frac{1}{aH^2} \frac{d^2a}{dt^2}, \\
		j(t) &= \frac{1}{aH^3} \frac{d^3a}{dt^3}, \\
		s(t) &= \frac{1}{aH^4} \frac{d^4a}{dt^4}, \\
		l(t) &= \frac{1}{aH^5} \frac{d^5a}{dt^5},\\
        m(t) &= \frac{1}{aH^6} \frac{d^6a}{dt^6},
	\end{aligned}
\end{equation}
where $H_0$, $q_0$, $j_0$, $s_0$, $l_0$ and $m_0$ are the cosmographic parameters at the present time, known as the Hubble constant, deceleration, jerk, snap,lerk and $m$ parameters, respectively.
As we can see in the above equations, cosmographic parameters are completely independent of DE models. 
The cosmographic parameters are extremely valuable observables for extracting information about the universe's expansion when calculated at the present time. Each parameter has a specific physical meaning, making them suitable for explaining the universe's expansion history. The Hubble function, $H$, indicates whether the universe is in an expansion ($\dot{a} > 0$) or contraction ($\dot{a} < 0$) phase. The sign of the deceleration parameter, $q$, determines whether the universe's expansion is accelerating or decelerating. When $q < 0$, $\ddot{a} > 0$, it reveals that the universe is in an accelerating phase. Other cosmographic parameters reveal more  important physical facts at higher redshifts. For more information and details about the physical meanings of cosmographic parameters, we refer the reader to \citep{Pourojaghi:2021den}.

Based on the cosmographic approach, we can expand the scale factor around the present time as follows \citep{Sahni2000}:
\begin{equation}\label{c2}
	\begin{aligned}
		a(t) \simeq &1 + H_0 (t - t_0) - \frac{q_0}{2!} H_0^2 (t - t_0)^2 + \frac{j_0}{3!} H_0^3 (t - t_0)^3 \\
		&+ \frac{s_0}{4!} H_0^4 (t - t_0)^4 + \frac{l_0}{5!} H_0^5 (t - t_0)^5 + \frac{m_0}{6!} H_0^6 (t - t_0)^6 + \cdots
	\end{aligned}
\end{equation}
Using this method, we can expand the Hubble parameter around the present time by employing the relationships between various time derivatives of the Hubble parameter and the cosmographic parameters.
It is worth noting that we are not limited to using a specific series in the cosmography approach. In fact, we can use different expansions such as Taylor series, Padé polynomials, Chebyshev approximations, logarithmic series, and others to reconstruct the Hubble parameter \citep{Capozziello:2018jya, Yang:2019vgk, 10.1093/mnras/staa871, Rezaei:2020lfy, Banerjee:2020bjq, Hu:2022udt, Capozziello:2017nbu, Capozziello:2020ctn}. The key point is that our expansion must not diverge at high redshifts and should minimize truncation errors.

As described in \citep{Pourojaghi:2022zrh}, the rational Padé approximation is a good choice for reconstructing the Hubble parameter. Since the rational Padé approximation decreases the divergence amplitude, it can expand the convergence domain of the approximation. Moreover, the Padé-cosmography works well at both high redshifts where GRBs and QSOs have been observed and at low redshift where we have observations from SNIa and BAO measurements \citep{Pourojaghi:2022zrh}. Based on Padé (3,2) cosmography, the dimensionless Hubble parameter $E(z)=H(z)/H_0$ can be reconstructed as follows:
\begin{equation}\label{eq:eq3}
E(z)=P_{3,2}(z) = \frac{P_0 + P_1 z + P_2 z^2 + P_3 z^3}{1 + Q_1 z + Q_2 z^2}\;,
\end{equation}
where
\begin{equation}\label{eq:eq4}
\begin{aligned}
P_0 =& 1;,\\
P_1=& E_{1,0} + Q_1;,\\
P_2=&  E_{1,0}Q_1 + \frac{E_{2,0}}{2} + Q_2;,\\
P_3=& E_{1,0}Q_2 + \frac{E_{2,0}Q_1}{2} + \frac{E_{3,0}}{6};,\\
Q_1=& \frac{-3E_{2,0}E_{5,0} + 5E_{3,0}E_{4,0}}{15E_{2,0}E_{4,0} - 20E_{3,0}^2};,\\
Q_2=& \frac{4E_{3,0}E_{5,0} - 5E_{4,0}^2} {60E_{2,0}E_{4,0} - 80E_{3,0}^2};,
\end{aligned}
\end{equation}
and $E_{n,0}$ represents $\frac{H_{n}\vert_{z=0}}{H_{0}}$ wherein $H_{n}\vert_{z=0}$ is calculated from
\begin{eqnarray}
\begin{aligned}
H_1\vert_{z=0}=& \frac{dH}{dz}\vert_{z=0}= H_{0}(1+q_0);,\\
H_2\vert_{z=0}=& \frac{d^{2}H}{dz^2}\vert_{z=0}=H_{0} (j_0-q_0^{2});,\\
H_3\vert_{z=0}=& \frac{d^{3}H}{dz^3}\vert_{z=0}=H_{0} (-4j_0q_0-3j_0+3q_0^{3} + 3q_0^{2} -s_0);,\\
H_4\vert_{z=0}=& \frac{d^{4}H}{dz^4}\vert_{z=0}= H_{0} (-4 j_0^{2} + 25 j_0 q_0^{2}+32j_0q_0 +12j_0\\
& +l_0- 15 q_0^{4}-24q_0^{3}-12q_0^{2}+7q_0s_0+8s_0);,\\
H_5\vert_{z=0}=& \frac{d^{5}H}{dz^5}\vert_{z=0}=H_{0} (70j_0^{2}q_0+60j_0^{2}-210j_0q_0^{3}\\
& -375j_0q_0^{2}-240j_0q_0+15j_0s_0-60j_0-11l_0q_0 \\
& -15l_0-m_0+105q_0^{5}+225 q_0^{4} + 180 q_0^{3} - 60 q_0^{2} s_0 \\
& + 60 q_0^{2} - 105 q_0 s_0 - 60 s_0);.
\end{aligned}
\end{eqnarray}\label{eq5}
The free parameters in this approach are $q_0$, $j_0$, $s_0$, $l_0$, and $m_0$ are determined utilizing observational data. Furthermore, we can calculate cosmographic parameters based on the model parameters by taking the derivative of the specific Hubble parameter in each cosmological model. For instance, the Hubble parameter in a flat-$\Lambda$CDM model is written as:
\begin{equation}\label{eq:c6}
	H(z) = H_0 \sqrt{\Omega_{m0}(1+z)^3 + (1-\Omega_{m0})}.
\end{equation}

So the cosmographic parameters in this model can be obtained based on the model parameter $\Omega_{m0}$ as follows \citep{Lusso:2019akb}:
\begin{equation}\label{c7}
	\begin{aligned}
		q_0 &=  \frac{3}{2} \Omega_{m0} - 1, \\
		j_0 &= 1, \\
		s_0 &= 1 - \frac{9}{2} \Omega_{m0}, \\
		l_0 &= 1 + 3 \Omega_{m0} + \frac{27}{2} \Omega_{m0}^2,\\
        m_0 &= -\frac{81}{4}\Omega_{m0}^3 - 81\Omega_{m0}^2 -\frac{27}{2}\Omega_{m0} +1.
	\end{aligned}
\end{equation}
In the flat-$\Lambda$CDM model, the jerk parameter $j_0$ is equal to $+1.0$ independent of $\Omega_{m0}$. In Appendix \ref{app:apx1}, we will show the robustness of the Padé(3,2) compared to other approximations such as Padé(2,2) and Taylor approximations for reconstructing the Hubble parameter and subsequently the luminosity distance.

\section{Observational data and $\chi^2$ function}{\label{sec:data}}
In this section, we briefly introduce the observational data used in our analysis and construct the $\chi^2$ function for each dataset. These datasets consist of the first year's baryon acoustic oscillation (BAO) measurements from DESI \citep{DESI:2024mwx}, Supernovae luminosity distance data from the Dark Energy Survey 5 Year release (DES-SN5YR) \citep{DES:2024tys}, and the Pantheon+ supernovae sample \citep{Scolnic:2021amr}.

\subsection{DESI BAO:}
This dataset includes 12 measurements spanning various redshifts from $0.1$ to $4.16$, effectively at $0.295\leq z \leq 2.33$ as presented in \citep{DESI:2024mwx}. In our analysis, we incorporate all the measurements that provide isotropic and anisotropic BAO measurements. Specifically, we utilize the two isotropic measurements and ten anisotropic measurements reported in DESI's first data release. These are crucial as they offer precise constraints on the comoving distance $D_M(z)$, the angle-average distance $D_V(z)$, and the Hubble distance $D_H(z)$ across the redshift range, enhancing our assessment of the standard cosmological model through the window of the cosmographic framework. In our approach, we first apply the cosmographic expansion of $H(z)$ using Eq. (\ref{eq:eq3}), where we utilize the Padé-cosmography method to expand the Hubble parameter. This expansion is directly used to derive expressions for $D_M(z)$, $D_V(z)$, and $D_A(z)$ as follows:

\textbf{- Comoving distance $d_M(z)$:}
	\begin{equation}\label{eq:dM}
		D_M(z) = \frac{c}{H_0}\int_0^z\frac{ dz'}{E(z')},  
	\end{equation}
where $c$ is speed of light, and $E(z) = H(z)/H_0$. This quantity is measured .\\

\textbf{- Hubble distance $D_H(z)$:}
	\begin{equation}\label{eq:dH}
		D_H(z) = \frac{c }{H(z)}.
	\end{equation}
    
\textbf{- Angle-averaged distance $D_V(z)$:}
	\begin{equation}
		D_V(z) = \left[z D_M^2(z) D_H(z)\right]^{1/3},  
	\end{equation}
which combines the transverse and radial distances, allowing a measure that optimizes sensitivity to both.
For each of these quantities, the cosmographic expansion provides a model-independent way to reconstruct the late-time cosmic expansion. By directly incorporating Eq. (\ref{eq:eq3}) into each distance formula, we measure the above geometrical distances in the context of the cosmographic approach. This method can support
the results outlined in \cite{DESI:2024mwx}, enabling a precise, model-independent assessment of cosmic expansion dynamics based on the DESI BAO measurements.\\
The chi-square function ($\chi^2$) for the DESI BAO data is formulated as follows:
	\begin{equation}\label{bao4}
		\chi^2_{DESI \; BAO} = \chi^2_{iso} + \chi^2_{aniso}.
	\end{equation}
where $\chi^2_{iso}$ is the chi-square function for isotropic data and $\chi^2_{aniso}$ is the chi-square function for anisotropic data as follow:
	\begin{equation}\label{bao5}
		\chi^2_{iso} = \sum^{n}_{i=1}\frac{[(\frac{D_V}{r_d})^{th}(z_i)-(\frac{D_V}{r_d})^{obs}(z_i)]^2}{\sigma_{i}^2}.
	\end{equation}

	\begin{equation}\label{bao6}
		\chi^2_{\text{aniso}} = Q \cdot C^{-1}_{BAO} \cdot Q^T. 
	\end{equation}
where ${C}^{-1}_{BAO}$ is the covariance matrix as follows:
\begin{equation}
	\scalebox{0.6}{
		$\mathbf{C}^{-1}_{BAO} = 
		\left( \begin{array}{cccccccccc}
			19.95 & 3.64 & 0 & 0 & 0 & 0 & 0 & 0 & 0 & 0 \\
			3.64 & 3.35 & 0 & 0 & 0 & 0 & 0 & 0 & 0 & 0 \\
			0 & 0 & 11.86 & 2.66 & 0 & 0 & 0 & 0 & 0 & 0 \\
			0 & 0 & 2.66 & 3.37 & 0 & 0 & 0 & 0 & 0 & 0 \\
			0 & 0 & 0 & 0 & 15.03 & 4.68 & 0 & 0 & 0 & 0 \\
			0 & 0 & 0 & 0 & 4.68 & 9.62 & 0 & 0 & 0 & 0 \\
			0 & 0 & 0 & 0 & 0 & 0 & 2.62 & 1.91 & 0 & 0 \\
			0 & 0 & 0 & 0 & 0 & 0 & 1.91 & 7.06 & 0 & 0 \\
			0 & 0 & 0 & 0 & 0 & 0 & 0 & 0 & 1.47 & 3.86 \\
			0 & 0 & 0 & 0 & 0 & 0 & 0 & 0 & 3.86 & 44.79
		\end{array}\right)$
	}
\end{equation}

and
	\begin{equation}\label{bao7}
		\small
		Q_i  = \left\{ \begin{array}{rc}
			(\frac{D_M}{r_d})^{\text{th}}(z_i) - (\frac{D_M}{r_d})^{\text{obs}}(z_i), & \text{for } D_M \text{ measurements}, \\
			(\frac{D_H}{r_d})^{\text{th}}(z_i) - (\frac{D_H}{r_d})^{\text{obs}}(z_i), & \text{for } D_H \text{ measurements}.
		\end{array}\right.
	\end{equation}

\subsection{Pantheon+ SNIa:}
The Pantheon+ sample represents the most recent and comprehensive collection of spectroscopically confirmed Type Ia supernovae (SNIa), spanning a redshift range from $z = 0$ to $z = 2.3$. This extensive dataset comprises 1701 supernova light curves, which have been critical in deriving cosmological parameters, particularly in the Pantheon SNIa and SH0ES distance-ladder analyses. Compared to the original Pantheon dataset, Pantheon+ includes an increased number of SNIa at lower redshifts, largely due to the addition of the LOSS1, LOSS2, SOUSA, and CNIa0.2 samples, as detailed in Table 1 of \cite{Scolnic:2021amr}.
The Pantheon+ sample integrates data from 18 distinct datasets, each meticulously calibrated for systematic uncertainties, making it an invaluable resource for refining measurements of the Hubble constant $H_0$ parameter and also other cosmological parameters. Notably, it includes 77 supernovae from galaxies that are also Cepheid hosts, within the low-redshift range of $0.00122 \leq z \leq 0.01682$. This overlap allows for a robust cross-calibration with Cepheid distances, helping to resolve the degeneracy between $H_0$ and the absolute magnitude $M$ of SNIa.
To fit our cosmology with the Pantheon+ dataset, we use the following chi-square function:
	\begin{equation}\label{eq:P1}
		\chi^2_{PantheonPlus} = Q^T . (C_{stat+sys})^{-1} . Q,
	\end{equation}
where $C_{stat+sys}$ represents the covariance matrix for the Pantheon+ sample, incorporating both statistical and systematical uncertainties as described in \cite{Brout:2022vxf}. The vector $Q$ is defined to account for the Cepheid-hosted supernovae, allowing for a consistent treatment of $H_0$ and $M$:
	\begin{equation}\label{eq:P2}
		Q_i =  \left\{ \begin{array}{rc}
			m_i - M - \mu_{i}, & i \in \text{Cepheid hosts} \\
			m_i - M - \mu_{\text{model}}(z_i), & \text{otherwise} \;
		\end{array}\right.
	\end{equation}
where $m_i$ denotes the apparent magnitude of the $i^{\text{th}}$ supernova, and $\mu_i \equiv m_i - M$ is the distance modulus. For non-cepheid host galaxies, the distance modulus $\mu_{model}(z_i)$ is calculated based on the model's predicted luminosity distance at redshift $z_i$ as follows:
        \begin{equation}\label{eq:P3}
		\mu_{model}(z_i) = 5 \log_{10}(D_L(z_i)) + 25,
	\end{equation}
where $D_L(z_i)$ is the luminosity distance at redshift $z_i$, given by:
        \begin{equation}\label{eq:P4}
		D_L(z_i) = \frac{c(1+z_i)}{H_0} \int_{0}^{z_i} \frac{dz'}{E(z')},
	\end{equation}
Here, $c$ is the speed of light, $H_0$ is the Hubble constant, and $E(z)$ is the dimensionless Hubble parameter, which is replaced by Eq. (\ref{eq:c6}) for the flat-$\Lambda$CDM model and Eq. (\ref{eq:eq3}) for the cosmography approach.

\subsection{DES-SN5YR:}
The DES-SN5YR sample \cite{DES:2024tys} represents the largest and deepest single-sample supernova survey to date, comprising 1829 supernovae across a redshift range of $0.01<z<1.2$. Collected over the first five years of the DE Survey Supernova Program (DES-SN), this extensive dataset provides crucial insights into the properties of DE and the cosmic expansion history. Among the 1829 supernovae, 1635 are genuine DES measurements spanning $0.10<z<1.13$, augmented with 194 low-redshift ($z < 0.1$) Type Ia supernovae from the CfA/CSP foundation sample. This combined sample benefits from strong high-redshift coverage, making it an exceptionally robust resource for cosmological analyses and enabling more precise constraints on the nature of DE.
To fit a given cosmological model to DES-SN5YR data, we employ the $\chi^2$ function
	\begin{equation}\label{DES1}
		\chi^2_{DESY5} = \Delta\mu_i C^{-1}_{ij}\Delta\mu_j,
	\end{equation}
where $\Delta \mu_i=\mu(z_i)-\mu_i$ is the difference between the model prediction at redshift $z_i$ and the observed distance modulus $\mu_i$ and $C^{-1}_{ij}$ is the 1829 $\times$ 1829 covariance matrix incorporating statistical and systematical uncertainties. The DES collaboration provide Hubble diagram redshifts, distance moduli and the corresponding distance moduli errors $\sigma_{\mu_i}$. In line with DES data usage guidelines, we also add the distance modulus error $\sigma^2_{\mu_i}$  to the diagonal of the covariance matrix to ensure accurate treatment of individual supernova uncertainties.

\section{Numerical Results}{\label{sec:data-analysis}}
In this section, our primary goal is to constrain cosmographic parameters using observational data explained in the previous section, and compare them with those obtained from the flat- $\Lambda$CDM cosmology. To achieve this, we employ Padé (3,2)-cosmography and various combinations of observational data introduced in Sec. \ref{sec:data}. Our method involves minimizing the $\chi^2$ function using the Markov Chain Monte Carlo (MCMC) algorithm.
A possible deviation between the values of cosmographic parameters in the two scenarios is calculated as follows:
\begin{equation}\label{r1}
	\Delta = \frac{|X_1 - X_2|}{\sqrt{\sigma^2_{X_1} + \sigma^2_{X_2}}},
\end{equation}
where $X_1$ and $X_2$ are the best-fit values of a cosmographic parameter in each scenario, and $\sigma^2_{X_1}$ and $\sigma^2_{X_2}$ are their error bars.

In this regard, we consider five different data samples:
\begin{enumerate}
	\item \textit{DESI BAO}
	\item \textit{DES-SN5YR}
	\item \textit{Pantheon Plus SNIa}
	\item \textit{DESI BAO + DES-SN5YR}
	\item \textit{DESI BAO + Pantheon Plus}
\end{enumerate}

First, we constrain the cosmological parameter $\Omega_{m0}$ in the flat-$\Lambda$CDM model for each data sample and use Eq. (\ref{c7}) to calculate cosmographic parameters in the $\Lambda$CDM model. Then, using Eq. (\ref{eq:eq3}), we independently constrain the cosmographic parameters for each data sample and compare the results with the $\Lambda$CDM model.

\begin{figure} 
	\centering
	\includegraphics[width=8cm]{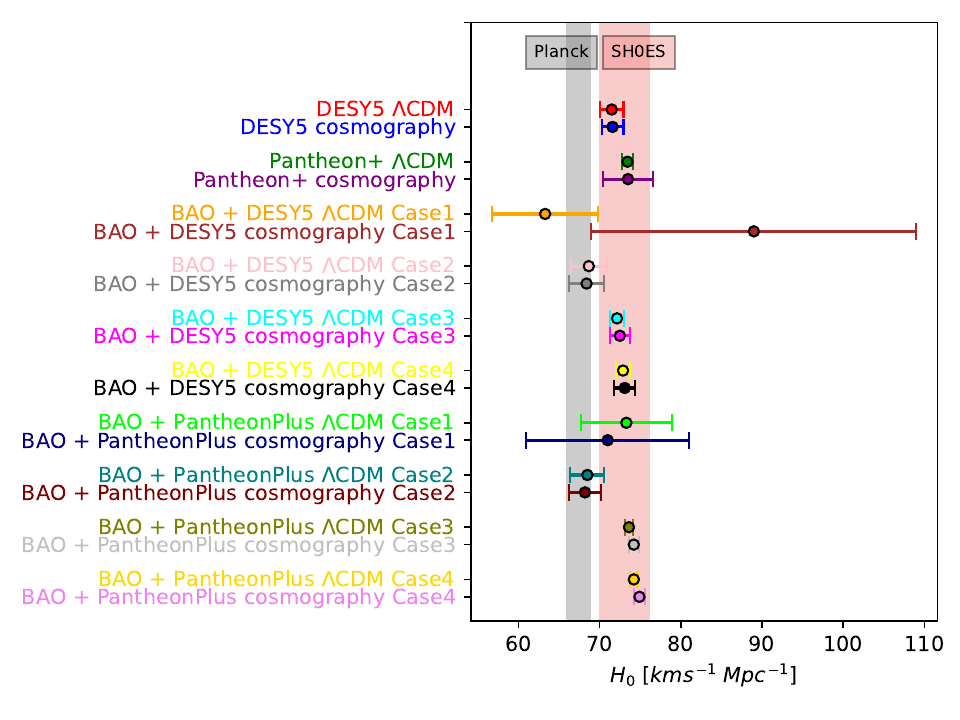}
	\caption{The values of $H_0$ parameters with their $3\sigma$ uncertainties, constrained in our analysis. Additionally, we show the $3\sigma$ values from Planck and SH0ES.}
	\label{fig_H0}
\end{figure}

\begin{table*}
\centering
\caption{The best-fit values of cosmological parameters with their $1\sigma$ uncertainty, obtained using various observational data in the flat-$\Lambda$CDM model (left), and the values of cosmographic parameters in the flat-$\Lambda$CDM model (right).}
\begin{tabular}{c c c c | c c c c c}
\hline \hline
Data  & $\Omega_{m0}$ & $H_0 r_d$ & $H_0$ & $q_0$ & $j_0$ & $s_0$ & $l_0$ & $m_0$\\
\hline
DESI BAO & $0.291^{+0.012}_{-0.014}$ & $102.4\pm 1.2$ & $-$ & $-0.564^{+0.018}_{-0.022}$ & $1.0$ & $-0.308^{+0.065}_{-0.055}$ & $3.01^{+0.13}_{-0.16}$ & $-10.28^{+0.98}_{-0.76}$\\
\hline 
DES-SN5YR & $0.337\pm 0.021$ & $-$ & $71.49\pm 0.46$ & $-0.494\pm 0.032$ & $1.0$ & $-0.518\pm 0.095$ & $3.55\pm 0.25$ & $-13.6\pm 1.6$\\
\hline
Pantheon+ & $0.334^{+0.016}_{-0.018}$ & $-$ & $73.44\pm 0.23$ & $-0.499^{+0.024}_{-0.028}$ & $1.0$ & $-0.503^{+0.083}_{-0.073}$ & $3.51^{+0.19}_{-0.23}$ & $-13.3^{+1.4}_{-1.1}$\\
\hline \hline		
\end{tabular}\label{Tab1}
\end{table*}

\begin{table*}
\centering
\caption{The best-fit values of cosmographic parameters with their $1\sigma$ uncertainty, obtained using various observational datasets in the Padé-cosmography. $\Delta_{q_0}$ and $\Delta_{j_0}$ indicate the deviations of the cosmographic parameters $q_0$ and $j_0$ from those values in $\Lambda$CDM cosmology, respectively.}
\begin{tabular}{c c c c c c c c c c c}
\hline \hline
Data  & $H_0 r_d$ & $H_0$ & $q_0$ & $j_0$ & $s_0$ & $l_0$ & $M$ & $\Delta_{q_0}$ & $\Delta_{j_0}$\\
\hline
DESI BAO & $103.9\pm 2.7$ & $-$ & $-0.669^{+0.088}_{-0.076}$ & $1.44^{+0.12}_{-0.20}$ & $0.34\pm 0.21$ & $2.80^{+0.85}_{-1.10}$ & $-14.05^{+0.46}_{-0.72}$ & $1.24 \sigma$ & $2.75 \sigma$\\
\hline
DES-SN5YR & $-$ & $71.64^{+0.46}_{-0.42}$ & $-0.503^{+0.043}_{-0.048}$ & $0.97\pm 0.17$ & $-0.56^{+0.20}_{-0.26}$ & $4.49\pm 0.78$ & $-11.9^{+3.2}_{-2.9}$ & $0.16 \sigma$ & $0.18 \sigma$\\
\hline
Pantheon+ & $-$ & $73.5^{+1.0}_{-1.2}$ & $-0.465\pm 0.036$ & $0.85^{+0.19}_{-0.12}$ & $-0.33^{+0.12}_{-0.29}$ & $1.7^{+2.1}_{-2.4}$ & $-7.8\pm 1.0$ & $0.77 \sigma$ & $0.97 \sigma$\\
\hline \hline		
\end{tabular}\label{Tab2}
\end{table*}

\begin{figure*} 
	\centering
	\includegraphics[width=6cm]{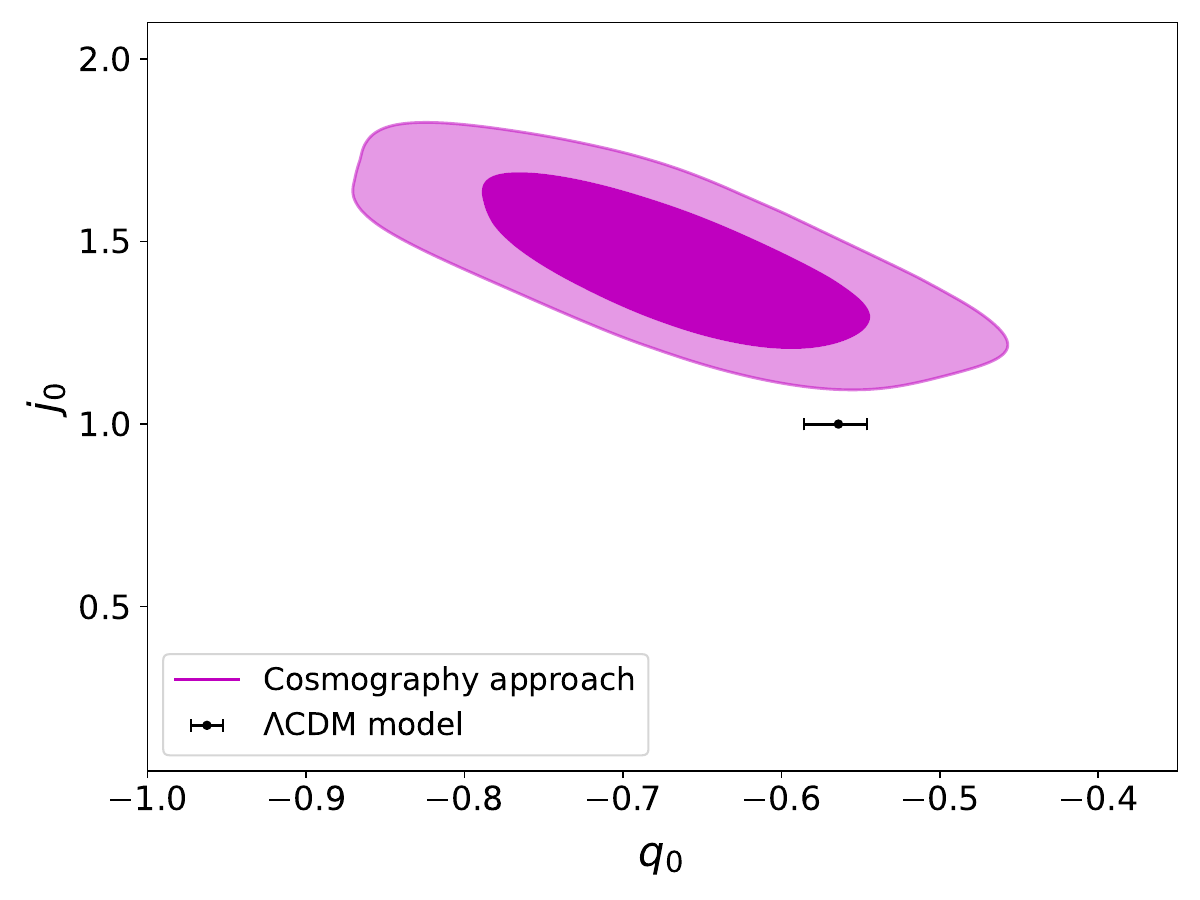}\includegraphics[width=6cm]{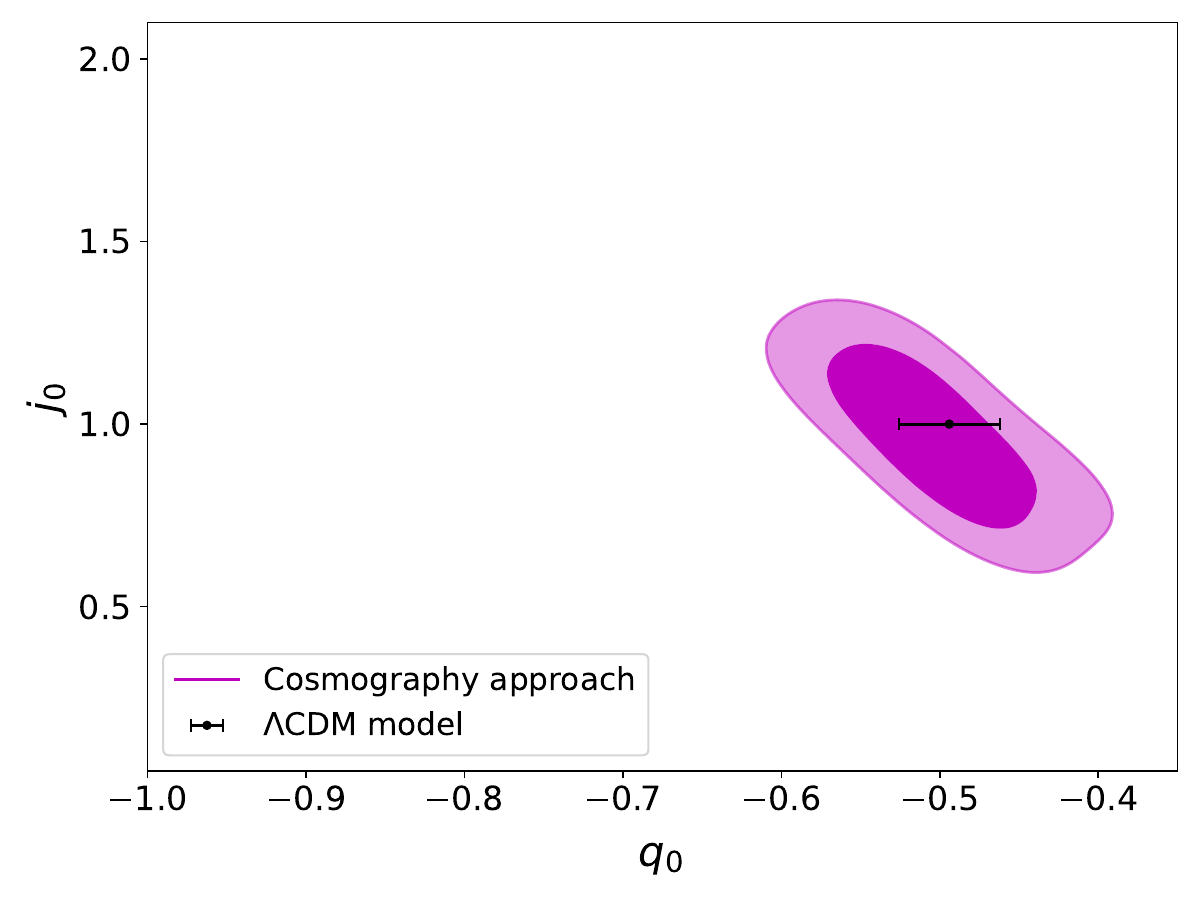}\includegraphics[width=6cm]{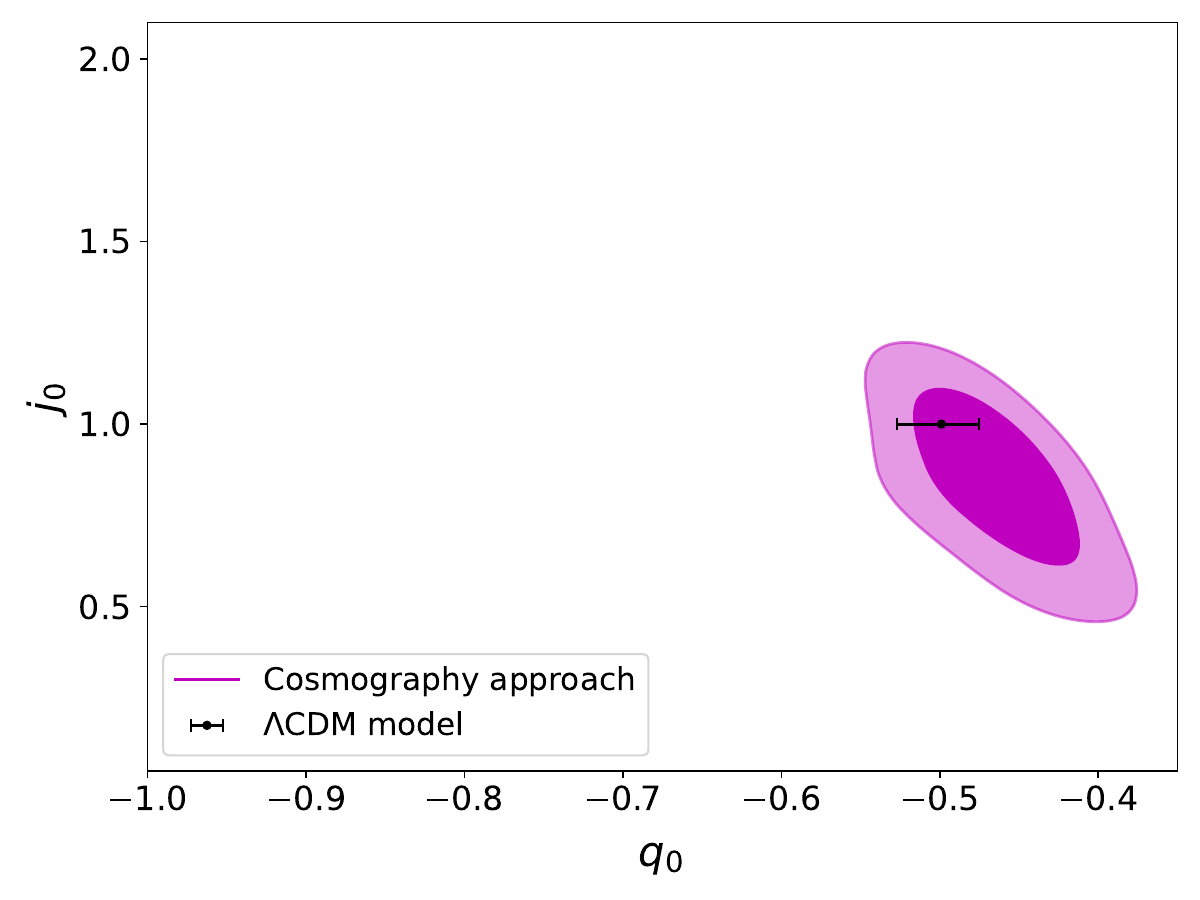}
	\caption{$1\sigma - 2\sigma$ confidence regions of the cosmographic parameters $q_0$ and $j_0$ obtained using the Padé-cosmography approach with the observational datasets of DESI BAO (left), DESY5 (middle), and Pantheon+ (right). The $\Lambda$CDM constraint for $q_0$ using the same datasets is shown for comparison.}
	\label{fig1}
\end{figure*}

\subsection{DESI BAO sample}
As we know, when using DESI BAO data alone, we cannot constrain $H_0$ and $r_d$ separately. Therefore, with this dataset, our free parameters in the flat-$\Lambda$CDM model are $\Omega_{m0}$ and $H_0 r_d$. In the cosmography approach, our free parameters are $H_0 r_d$, $q_0$, $j_0$, $s_0$, $l_0$ and $m_0$. Our results for the flat-$\Lambda$CDM model are presented in Table \ref{Tab1}. The first row shows the results for DESI BAO observations. Additionally, within the Padé(3,2)-cosmography, we present our results for DESI BAO observations in the first row of Table \ref{Tab2}. Comparing both results shows that DESI BAO alone confirms the flat-$\Lambda$CDM within the cosmography approach. In the left panel of Fig. \ref{fig1}, we show this consistency between the flat-$\Lambda$CDM model and the Padé(3,2)-cosmographic approach in the $q_0-j_0$ plane. Using Eq. (\ref{r1}), we obtain the deviation between both scenarios as $\Delta_{q_0}=1.24\sigma$ and $\Delta_{j_0}=2.75\sigma$ for $q_0$ and $j_0$, respectively. Both values are under $3 \sigma$, supporting the consistency between the model and cosmography.
Moreover, the values of $H_0 r_d$ in both scenarios, with a small deviation of $0.51\sigma$, are in full agreement with each other (see the values of $H_0 r_d$ in the first rows of Tables \ref{Tab1} and \ref{Tab2}). This confirmation of the standard $\Lambda$CDM model obtained in the cosmography method is consistent with previous results obtained from DESI BAO measurements \citep{DESI:2024mwx, Pourojaghi:2024tmw, Giare:2024gpk}.

\subsection{DES-SN5YR sample}
When using only the SNIa sample, the free parameters in the flat-$\Lambda$CDM model are $\Omega_{m0}$, $H_0$, and $M$. Since the absolute magnitude $M$ and the Hubble constant $H_0$ are degenerate, we cannot constrain both simultaneously. Therefore, we must fix one and constrain the other when using observational data from SNIa compilation alone. In this work, we apply a tight prior from Cepheid observations on the absolute magnitude as $M = -19.253 \pm 0.027$ \citep{Perivolaropoulos:2022khd}. Consequently, in the flat-$\Lambda$CDM model, we have just two free parameters: $\Omega_{m0}$ and $H_0$. In the cosmography method, we have five free parameters: $H_0$, $q_0$, $j_0$, $s_0$, $l_0$ and $m_0$. Our numerical results are presented in the second rows of Tables \ref{Tab1} and \ref{Tab2} for the flat-$\Lambda$CDM model and the Padé-cosmography approach, respectively. The results show that the best-fit values of the cosmographic parameters in the flat-$\Lambda$CDM and cosmography approaches are fully consistent with each other. This result agrees with the original DES paper \citep{DES:2024tys}, where they showed support for a constant EoS $w_{\Lambda} = -1$ in the context of the $w_0w_a$ parametrizations. For the first two cosmographic parameters, we refer to the middle panel of Fig. \ref{fig1}. We obtain $\Delta_{q_0}=0.16\sigma$ and $\Delta_{j_0}=0.18\sigma$ differences, respectively, between the $q_0$ and $j_0$ parameters of both scenarios, supporting the consistency between them.
Moreover, the values of the Hubble constant in the flat-$\Lambda$CDM model ($H_0 = 71.49 \pm 0.46$ km s$^{-1}$ Mpc$^{-1}$) and the Padé-cosmography approach ($H_0 = 71.64^{+0.46}_{-0.42}$ km s$^{-1}$ Mpc$^{-1}$) are in full agreement with a small difference of $0.2\sigma$, and both are consistent with the SH0ES value \citep{Riess:2021jrx} (see Fig. \ref{fig_H0}).

\subsection{Pantheon+ SNIa}
In the case of the Pantheon+ sample, to eliminate the degeneracy between $H_0$ and $M$, we put prior on $M$ from Cepheid host SNIa and then constrain the other free parameters. Our numerical results for the flat- $\Lambda$CDM model are presented in the third row of Table \ref{Tab1}, and for Padé-cosmography, they are presented in the third row of Table \ref{Tab2}. Within $1\sigma$ uncertainty, we observe consistency between the two scenarios, as shown in the right panel of Fig. \ref{fig1}. Quantitatively, we obtain $\Delta_{q_0}=0.77\sigma$ and $\Delta_{j_0}=0.97\sigma$ differences, respectively, for the $q_0$ and $j_0$ parameters of the flat-$\Lambda$CDM and Padé-cosmography approaches. Moreover, the best-fit values of $H_0$ in both scenarios agree with each other, with a small difference of $0.05\sigma$. Both values are within the SH0ES region at the $1\sigma$ level (see Fig. \ref{fig_H0}). These results for $H_0$ are in agreement with the findings of the original work \citep{Scolnic:2021amr}, which supported the standard flat-$\Lambda$CDM model within the general contexts of $w$CDM and $w_0w_a$ parametrizations.

\begin{table*}
\centering
\caption{The best-fit values of cosmological parameters with their $1\sigma$ uncertainty, obtained using DESI BAO + DES-SN5YR sample in the flat-$\Lambda$CDM model (left), and the values of cosmographic parameters in the flat-$\Lambda$CDM model (right).}
\begin{tabular}{c c c c| c c c c c}
\hline \hline
Case  & $\Omega_{m0}$ & $H_0$ & $M$ & $q_0$ & $j_0$ & $s_0$ & $l_0$ & $m_0$\\
\hline
Case 1 & $0.306\pm 0.012$ & $63.3^{+1.9}{-3.5}$ & $-19.539^{+0.069}{-0.12}$ & $-0.542\pm 0.018$ & $1$ & $-0.375\pm 0.053$ & $3.18\pm 0.13$ & $-11.28\pm 0.82$\\
\hline
Case 2 & $0.305\pm 0.012$ & $68.68\pm 0.69$ & $-19.360\pm 0.016$ & $-0.543\pm 0.018$ & $1$ & $-0.371\pm 0.053$ & $3.17\pm 0.13$ & $-11.22\pm 0.80$\\
\hline
Case 3 & $0.305^{+0.011}{-0.012}$ & $72.15\pm 0.30$ & $-$ & $-0.543^{+0.016}{-0.019}$ & $1$ & $-0.371^{+0.056}{-0.049}$ & $3.17^{+0.12}{-0.14}$ & $-11.21^{+0.88}{-0.71}$\\
\hline
Case 4 & $0.2540^{+0.0067}{-0.0075}$ & $72.89\pm 0.28$ & $-$ & $-0.619^{+0.010}{-0.011}$ & $1$ & $-0.143^{+0.034}{-0.030}$ & $2.634^{+0.064}{-0.075}$ & $-7.99^{+0.44}{-0.38}$\\
\hline \hline		
\end{tabular}\label{Tab3}
\end{table*}

\begin{table*}
\centering
\caption{The best-fit values of cosmographic parameters with their $1\sigma$ uncertainty, obtained using DESI BAO + DES-SN5YR sample in the Padé-cosmographic approach. $\Delta_{q_0}$ and $\Delta_{j_0}$ indicate the deviations of the cosmographic parameters $q_0$ and $j_0$ from those values in $\Lambda$CDM cosmology, respectively.}
\footnotesize
\begin{tabular}{c c c c c c c c c c}
\hline \hline
Case & $H_0$ & $M$ & $q_0$ & $j_0$ & $s_0$ & $l_0$ & $m_0$ & $\Delta_{q_0}$ & $\Delta_{j_0}$\\
\hline
Case 1 & $89.0\pm 9.5$ & $-18.81^{+0.31}_{-0.34}$ & $-0.575\pm 0.044$ & $1.34^{+0.21}_{-0.28}$ & $-0.358^{+0.081}_{-0.140}$ & $1.71^{+0.59}_{-0.97}$ & $-9.48^{+0.94}_{-1.50}$ & $0.69 \sigma$ & $1.36 \sigma$\\
\hline
Case 2 & $68.44\pm 0.79$ & $-19.361\pm 0.018$ & $-0.525^{+0.061}_{-0.054}$ & $0.96^{+0.28}_{-0.36}$ & $-0.94\pm 0.20$ & $1.81\pm 0.71$ & $-9.52^{+0.62}_{-1.00}$ & $0.03 \sigma$ & $0.13 \sigma$\\
\hline
Case 3 & $72.48^{+0.48}_{-0.44}$ & $-$ & $-0.472\pm 0.041$ & $0.785^{+0.053}_{-0.160}$ & $-0.492^{+0.190}_{-0.070}$ & $2.79\pm 0.49$ & $-12.55\pm 0.96$ & $1.59 \sigma$ & $1.35 \sigma$\\
\hline
Case 4 & $73.10\pm 0.47$ & $-$ & $-0.495^{+0.054}_{-0.041}$ & $0.40^{+0.10}_{-0.17}$ & $-0.509^{+0.080}_{-0.19}$ & $0.71^{+0.59}_{-0.83}$ & $-13.31^{+0.40}_{-1.10}$ & $2.58 \sigma$ & $4.62 \sigma$\\
\hline \hline		
\end{tabular}\label{Tab4}
\end{table*}

\begin{figure*} 
	\centering
	\includegraphics[width=6cm]{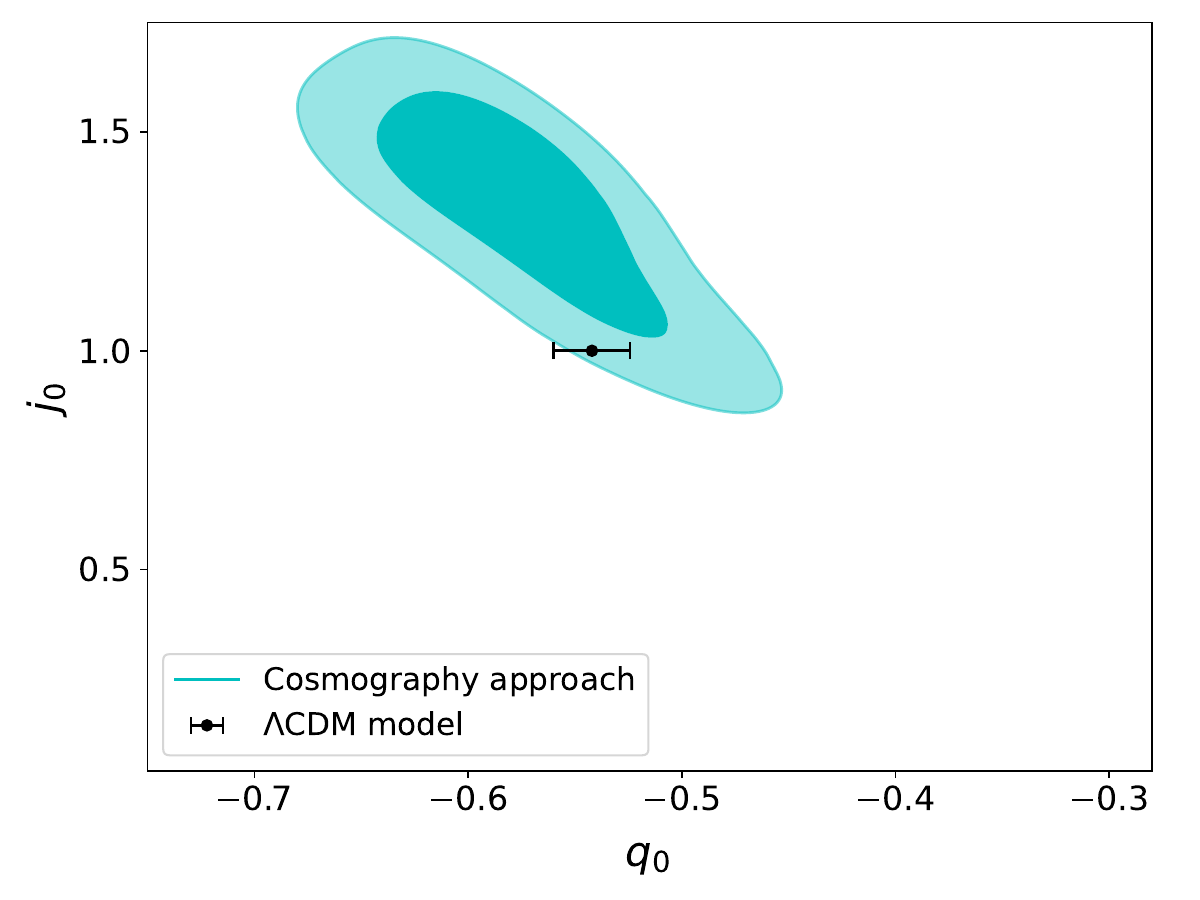}\includegraphics[width=6cm]{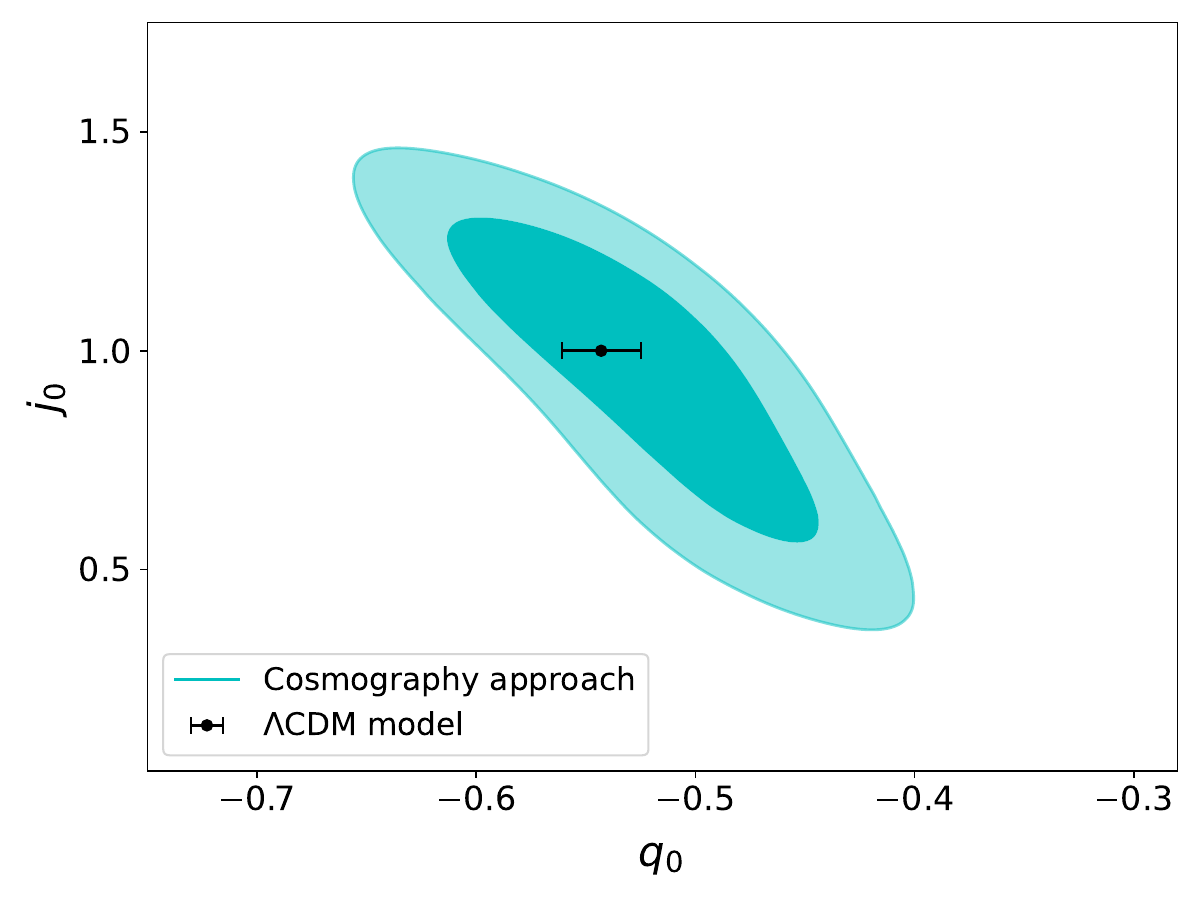}
	\includegraphics[width=6cm]{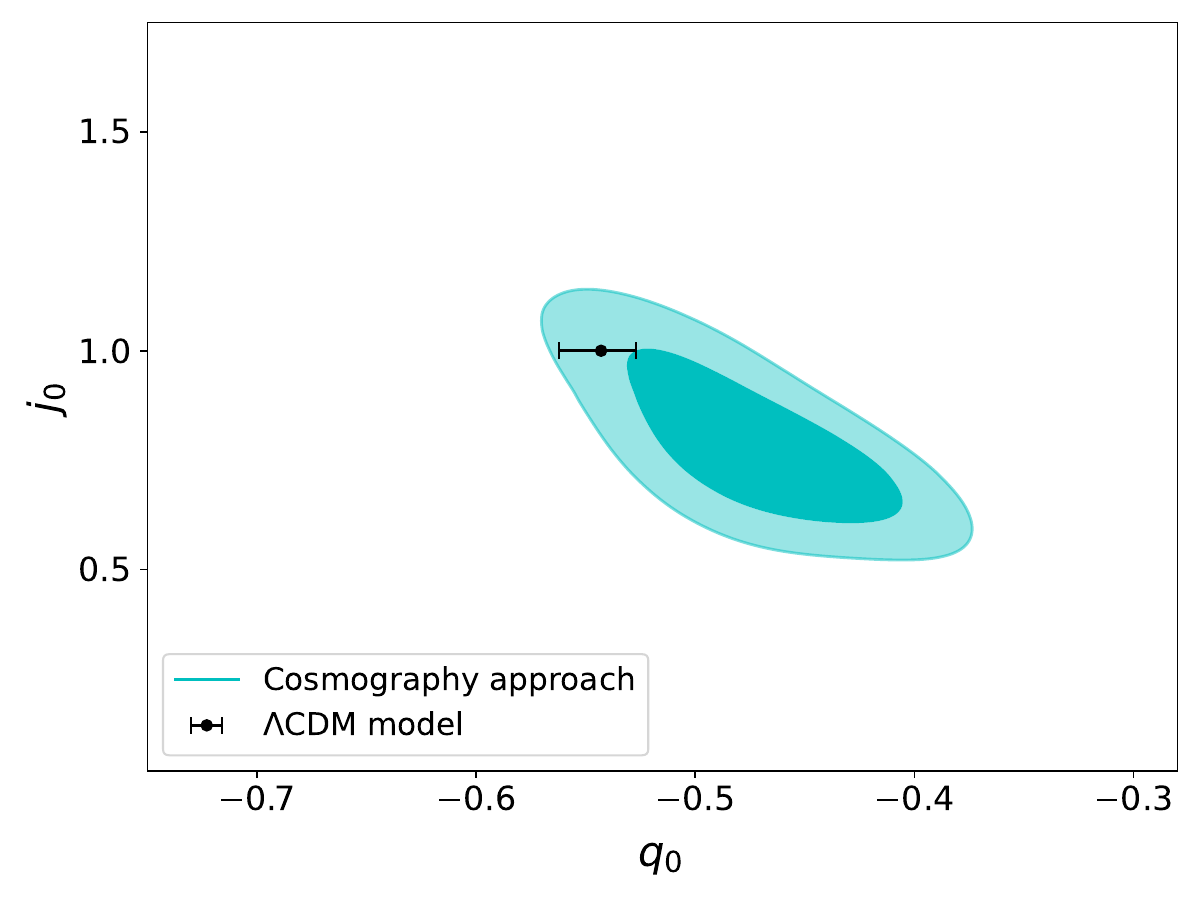}\includegraphics[width=6cm]{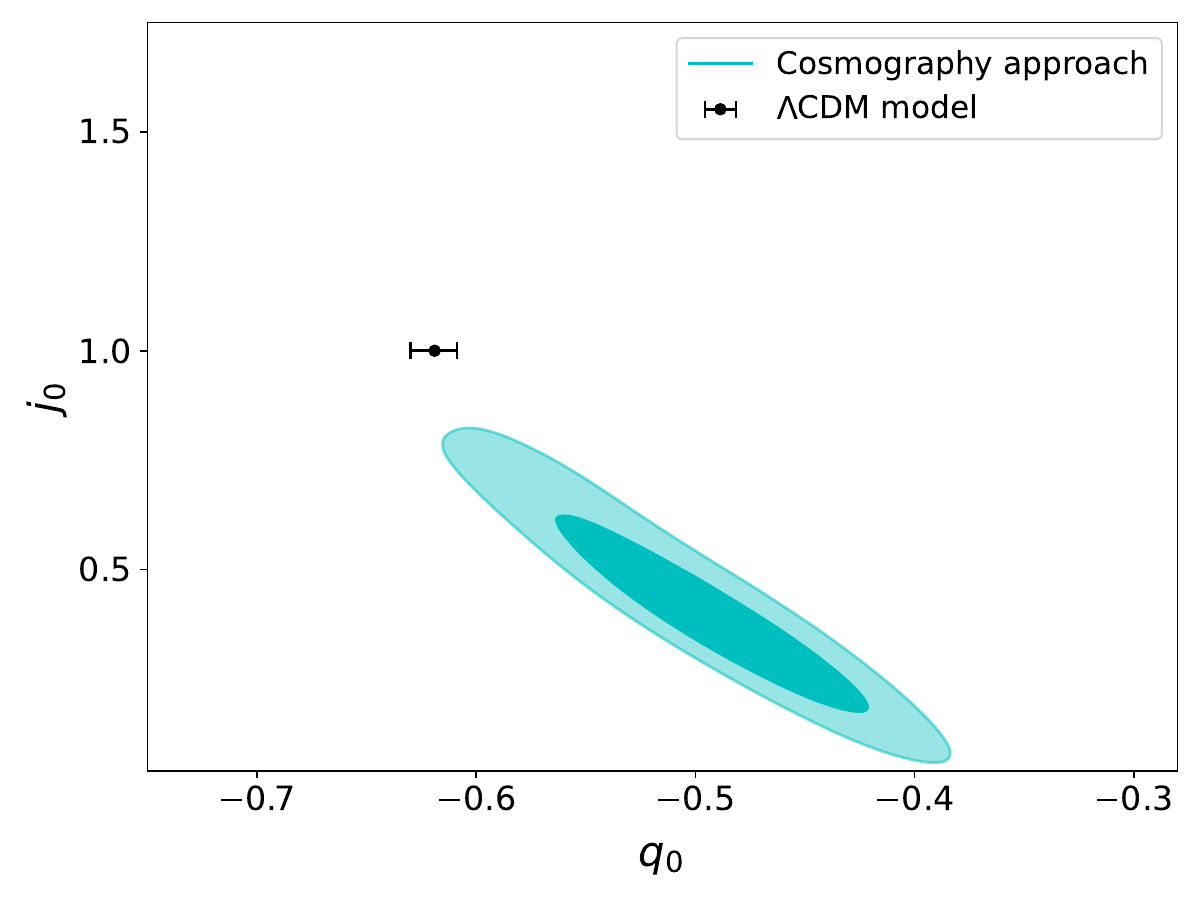}
	\caption{$1\sigma - 2\sigma$ confidence regions of the cosmographic parameters $q_0$ and $j_0$ obtained using the Padé-cosmographic approach with DESI BAO + DES-SN5YR sample. The upper left (right) panel shows Case 1 (Case 2) and lower left (right) indicate Case 3 (Case 4).}
	\label{fig2}
\end{figure*}

\subsection{DESI BAO + DES-SN5YR sample}{\label{SubSec:BAO_DES}}
In this section, we reanalyze the combination of datasets DESI BAO and DES-SN5YR. This combination allows us to constrain $r_d$ and $H_0$ separately. Moreover, using this combination, we can break the degeneracy between $H_0$ and $M$ without a need to prior on $M$. In the $\Lambda$CDM model, our free parameters are $\Omega_{m0}$, $H_0$, $M$, and $r_d$. In the cosmography approach, the free parameters are $H_0$, $M$, $r_d$, $q_0$, $j_0$, $s_0$, $l_0$ and $m_0$. As we know, when the number of free parameters increases, the uncertainty on parameters becomes large. For the absolute magnitude $M$, we can consider two cases: firstly, we can use the prior $M = -19.253 \pm 0.027$ \citep{Perivolaropoulos:2022khd}, meaning that we utilize a prior from Cepheid observations. It should be noted that Cepheid variables are not directly used for the calibration of Type Ia supernovae in the DES-SN5YR sample. Instead, the calibration of SNIa within this survey typically relies on previously established calibrations performed by other surveys that have used Cepheid variables, such as those bridging nearby supernovae with precise distance measurements. Secondly, we can consider $M$ as a free parameter, allowing it to be determined by other observations. In the latter case, we aim to calibrate the absolute magnitude $M$ of SNIa using the DESI BAO observations at higher redshifts than Cepheids. We extend our analysis by considering the same idea for the sound horizon $r_d$. Thus, we first consider the Planck prior and in the second case, we allow it to be free. In the latter case, we determine $r_d$ using the combination of DESI-BAO+SNIa datasets. Therefore, in general, we set up our analysis based on four different cases for cosmological priors as follows:
\begin{itemize}
	\item \textit{Case 1}: We relax the priors on $M$ and $r_d$, allowing both to be free parameters.
	\item \textit{Case 2}: Planck prior is applied on the sound horizon ($r_d = 147.46 \pm 0.28$) \citep{DES:2024ywx} and $M$ is free.
	\item \textit{Case 3}: Cepheid prior is applied on the absolute magnitude ($M = -19.253 \pm 0.027$) \citep{Perivolaropoulos:2022khd} and $r_d$ is free.
	\item \textit{Case 4}: Both Planck and Cepheid priors are applied on $r_d$ and $M$, respectively.
\end{itemize}

The results of our analysis in both the flat-$\Lambda$CDM and cosmography approaches for each case are as follows:

\textbf{Case 1:} As reported in the first rows of Tables \ref{Tab3} and \ref{Tab4}, the best-fit values of $H_0$ in the flat-$\Lambda$CDM model and the Padé-cosmography approach show a significant deviation. However, considering their error bars, these values meet each other with a $2.6\sigma$ deviation. Moreover, in this case for the flat-$\Lambda$CDM model, we obtain $\Omega_{m0} = 0.306 \pm 0.012$, so we get $q_0 = -0.542 \pm 0.018$ based on Eq. (\ref{c7}). As shown in the upper left panel of Fig. \ref{fig2}, this value is completely within the \bm{$1\sigma$ }region ($\Delta_{q_0}=0.69\sigma$ difference) of its value in the Padé-cosmographic method. Additionally, the jerk parameter, $j_0$, in the $\Lambda$CDM model is located well within the \bm{$2\sigma$} confidence region ($\Delta_{j_0}=1.36\sigma$ difference) of the cosmographic method, supporting the idea that the $\Lambda$CDM and cosmography approaches agree with each other. In this case, without the Cepheid prior on $M$ and the Planck prior on $r_d$, our analysis (see Fig. \ref{fig_H0}) results in larger uncertainties on $H_0$. Consequently, in both the $\Lambda$CDM and cosmography methods, we can reconcile the Planck value \citep{Planck:2018vyg} and the Cepheid-based observed value \citep{Riess:2021jrx} for $H_0$.

\textbf{Case 2:} In this case, and in the flat-$\Lambda$CDM model, the best-fit value of $\Omega_{m0}$ did not change compared to the previous case. However, we obtained a larger value for $H_0$, which is fully consistent with the value of $H_0$ constrained in the cosmography method with a small difference of $0.23\sigma$ (see Tables \ref{Tab3} and \ref{Tab4}). As we can see in the upper right panel of Fig. \ref{fig2}, the difference between the values of the cosmographic parameters $q_0$ and $j_0$ in the flat-$\Lambda$CDM model and the Padé-cosmographic approach decreases compared to Case 1. This difference is $\Delta_{q_0}=0.03\sigma$ for the $q_0$ parameter and $\Delta_{j_0}=0.13\sigma$ for the $j_0$ parameter. Our constraints on $H_0$ for both the flat-$\Lambda$CDM and Padé-cosmography are shown in Fig. \ref{fig_H0}. Both scenarios support the Planck value of $H_0$ \citep{Planck:2018vyg}, and our result for Padé-cosmography is in agreement with the results in \citep{DES:2024ywx}. The support of the Planck range in our analysis is due to imposing the Planck prior on $r_d$ while $M$ is free.

\textbf{Case 3:} As can be seen in Table \ref{Tab3}, in this case, the best-fit value of $\Omega_{m0}$ and, accordingly, $q_0$ has not changed significantly compared to the two previous cases. However, the value of the Hubble constant has increased from a value close to the Planck measurement ($H_0 = 67.4 \pm 0.5$ km s$^{-1}$ Mpc$^{-1}$) \citep{Planck:2018vyg} to a value close to the SH0ES measurement ($73.2 \pm 1.3$ km s$^{-1}$ Mpc$^{-1}$) \citep{Riess:2021jrx} compared to Case 2 (see Fig. \ref{fig_H0}) and the corresponding values for $H_0$ in the third rows of Tables \ref{Tab3} and \ref{Tab4}. This result was somewhat predictable because, in this case, we have fixed the value of $M$ with Cepheid prior, while $r_d$ is free. In the cosmography analysis, in the lower left panel of Fig. \ref{fig2} (see also the corresponding values in the third rows of Tables \ref{Tab3} and \ref{Tab4}), we clearly observe that the values of the cosmographic parameters in the $q_0-j_0$ plane in both the $\Lambda$CDM model and the cosmography approach are in agreement with each other. Quantitatively, we obtain $\Delta_{q_0}=1.59\sigma$ and $\Delta_{j_0}=1.35\sigma$ deviations between the flat-$\Lambda$CDM and Padé-cosmographic approaches for $q_0$ and $j_0$, respectively.

\textbf{Case 4:} In this case, we obtained $H_0 = 72.89 \pm 0.28$ km s$^{-1}$ Mpc$^{-1}$ in the flat-$\Lambda$CDM model and $H_0 = 73.10 \pm 0.47$ km s$^{-1}$ Mpc$^{-1}$ in the Padé-cosmographic approach, confirming each other with a small difference of \bm{$0.38\sigma$} and in full agreement with the SH0ES value (see Fig. \ref{fig_H0}). The last row of Table \ref{Tab3} shows that, in this case, the $\Lambda$CDM model yields a lower value for $\Omega_{m0}$ compared to other cases. According to Eq. (\ref{c7}), we get lower values for the cosmographic parameters in the context of the $\Lambda$CDM case. Here, we observe significant differences between the cosmographic parameters of the flat-$\Lambda$CDM model and those of the Padé-cosmography approach. In the lower-left panel of Fig. \ref{fig2}, we show a $\Delta_{q_0}=2.58\sigma$ deviation for $q_0$ values and an $\Delta_{j_0}=4.62\sigma$ deviation for $j_0$ values, indicating significant tension between the flat-$\Lambda$CDM model and Padé-cosmography.

\begin{table*}
\centering
\caption{Same as Table (\ref{Tab3}), but for DESI BAO + Pantheon Plus sample.}
\begin{tabular}{c c c c | c c c c c}
\hline \hline
Case  & $\Omega_{m0}$ & $H_0$ & $M$ & $q_0$ & $j_0$ & $s_0$ & $l_0$ & $m_0$\\
\hline
Case 1 & $0.309^{+0.011}_{-0.012}$ & $73.3^{+2.1}_{-3.5}$ & $-19.263^{+0.065}_{-0.100}$ & $-0.537^{+0.016}_{-0.019}$ & $1$ & $-0.389^{+0.056}_{-0.047}$ & $3.21^{+0.12}_{-0.14}$ & $-11.50^{+0.89}_{-0.70}$\\
\hline
Case 2 & $0.308^{+0.011}_{-0.013}$ & $68.49\pm 0.74$ & $-19.410\pm 0.02$ & $-0.538^{+0.017}_{-0.019}$ & $1$ & $-0.385^{+0.057}_{-0.051}$ & $3.20^{+0.12}_{-0.15}$ & $-11.43^{+0.90}_{-0.76}$\\
\hline
Case 3 & $0.309\pm 0.012$ & $73.62\pm 0.18$ & $-$ & $-0.536\pm 0.018$ & $1$ & $-0.391\pm 0.054$ & $3.22^{+0.13}_{-0.14}$ & $-11.52^{+0.88}_{-0.77}$\\
\hline
Case 4 & $0.2376\pm 0.0054$ & $74.22\pm 0.16$ & $-$ & $-0.6436\pm 0.0081$ & $1$ & $-0.069\pm 0.024$ & $2.475\pm 0.051$ & $-7.05\pm 0.30$\\
\hline \hline		
\end{tabular}\label{Tab5}
\end{table*}

\begin{table*}
\centering
\caption{Same as Table (\ref{Tab4}), but for DESI BAO + Pantheon Plus sample.}
\begin{tabular}{c c c c c c c c c c}
\hline \hline
Case & $H_0$ & $M$ & $q_0$ & $j_0$ & $s_0$ & $l_0$ & $m_0$ & $\Delta_{q_0}$ & $\Delta_{j_0}$\\
\hline
Case 1 & $71.2^{+5.5}_{-4.9}$ & $-19.32^{+0.18}_{-0.13}$ & $-0.457^{+0.033}_{-0.043}$ & $0.76\pm 0.17$ & $-0.145\pm 0.086$ & $3.08\pm 0.55$ & $-10.8\pm 1.1$ & $1.91 \sigma$ & $1.41 \sigma$\\
\hline
Case 2 & $68.19\pm 0.74$ & $-19.415\pm 0.02$ & $-0.524^{+0.021}_{-0.027}$ & $1.156^{+0.091}_{-0.035}$ & $0.14^{+0.25}_{-0.28}$ & $3.12\pm 0.79$ & $-9.03^{+0.63}_{-1.10}$ & $0.47 \sigma$ & $2.47 \sigma$\\
\hline
Case 3 & $74.22\pm 0.24$ & $-$ & $-0.472^{+0.041}_{-0.046}$ & $0.79^{+0.21}_{-0.17}$ & $-0.461^{+0.054}_{-0.110}$ & $2.60^{+0.69}_{-0.47}$ & $-13.02^{+0.91}_{-1.90}$ & $1.36 \sigma$ & $1.17 \sigma$\\
\hline
Case 4 & $74.91^{+0.26}_{-0.20}$ & $-$ & $-0.556^{+0.021}_{-0.037}$ & $0.456^{+0.082}_{-0.120}$ & $-0.452^{+0.230}_{-0.051}$ & $1.42^{+0.56}_{-0.43}$ & $-9.29^{+0.95}_{-0.50}$ & $2.91 \sigma$ & $4.95 \sigma$\\
\hline \hline	
\end{tabular}\label{Tab6}
\end{table*}

\begin{figure*} 
	\centering
	\includegraphics[width=6cm]{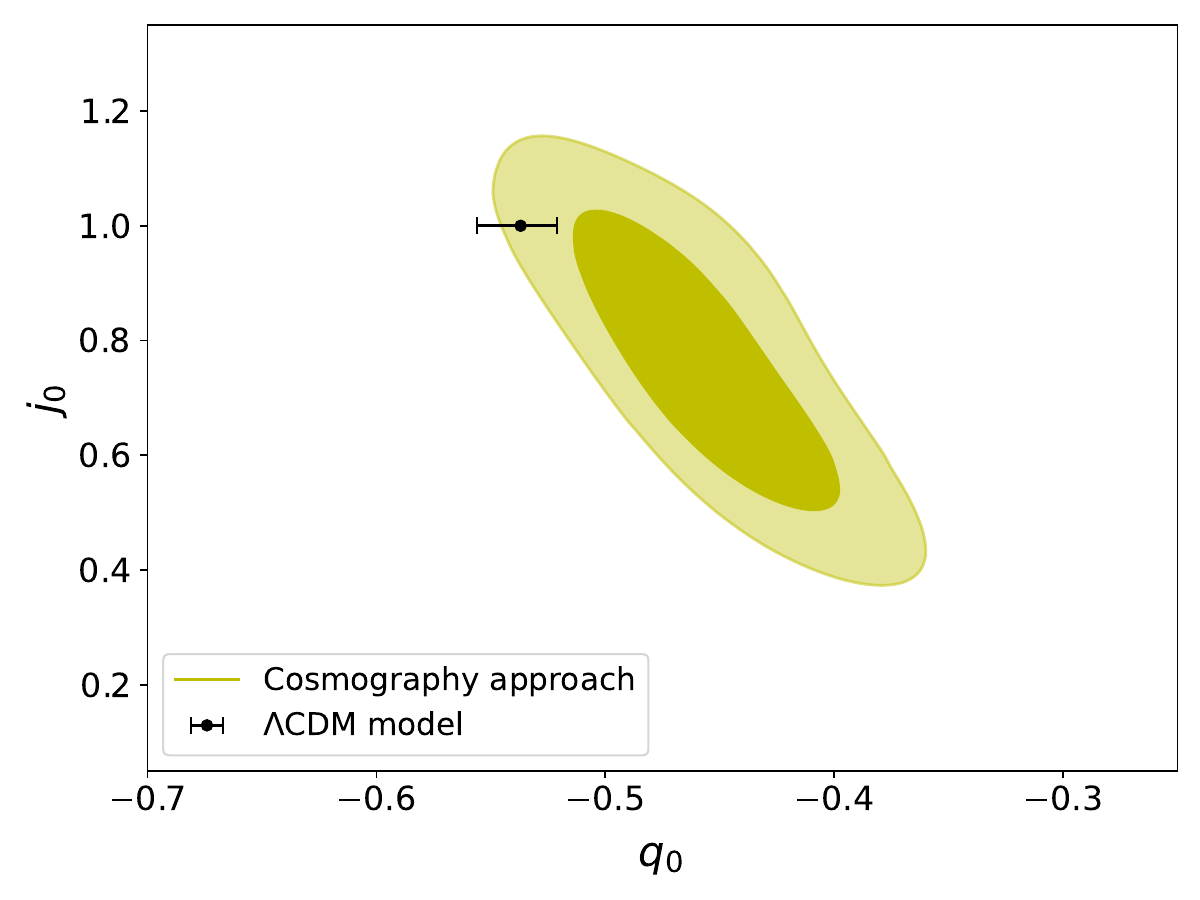}\includegraphics[width=6cm]{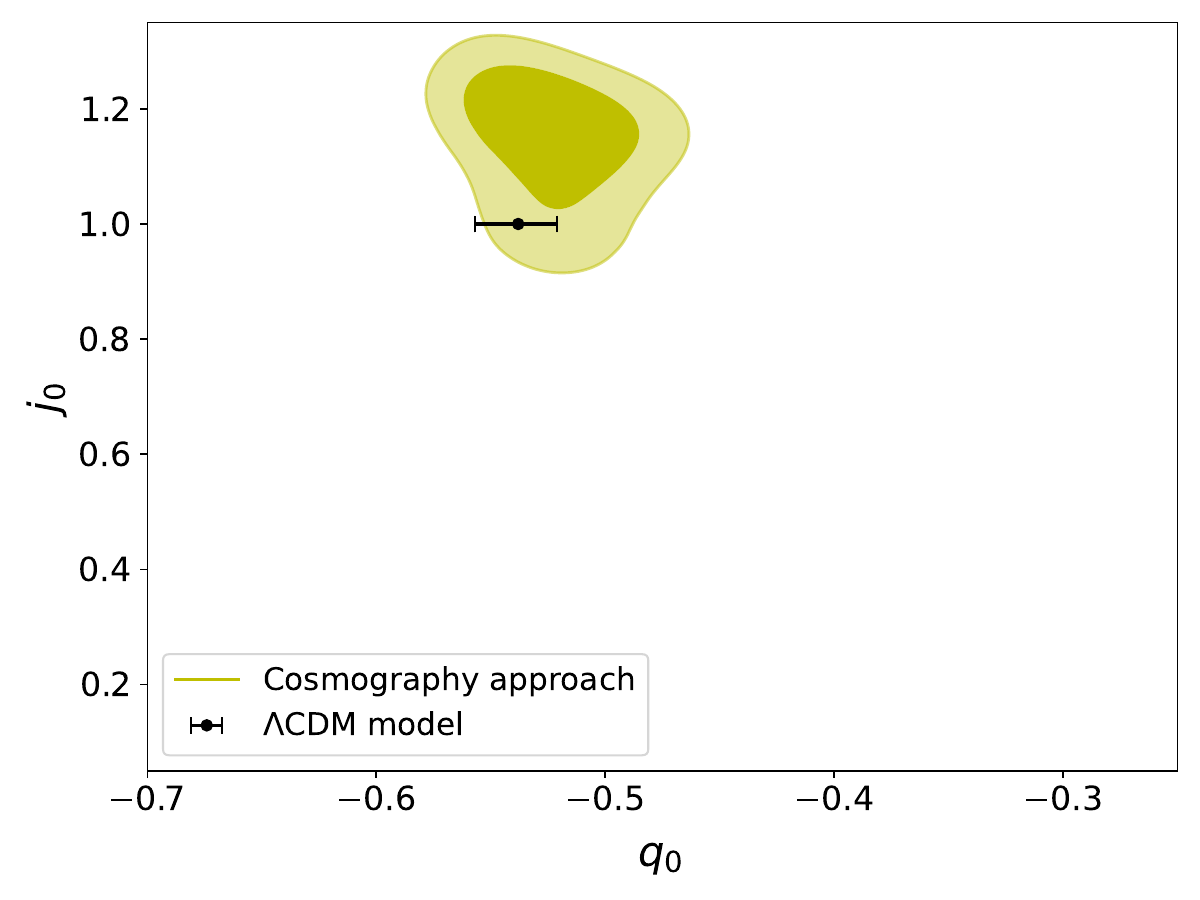}
	\includegraphics[width=6cm]{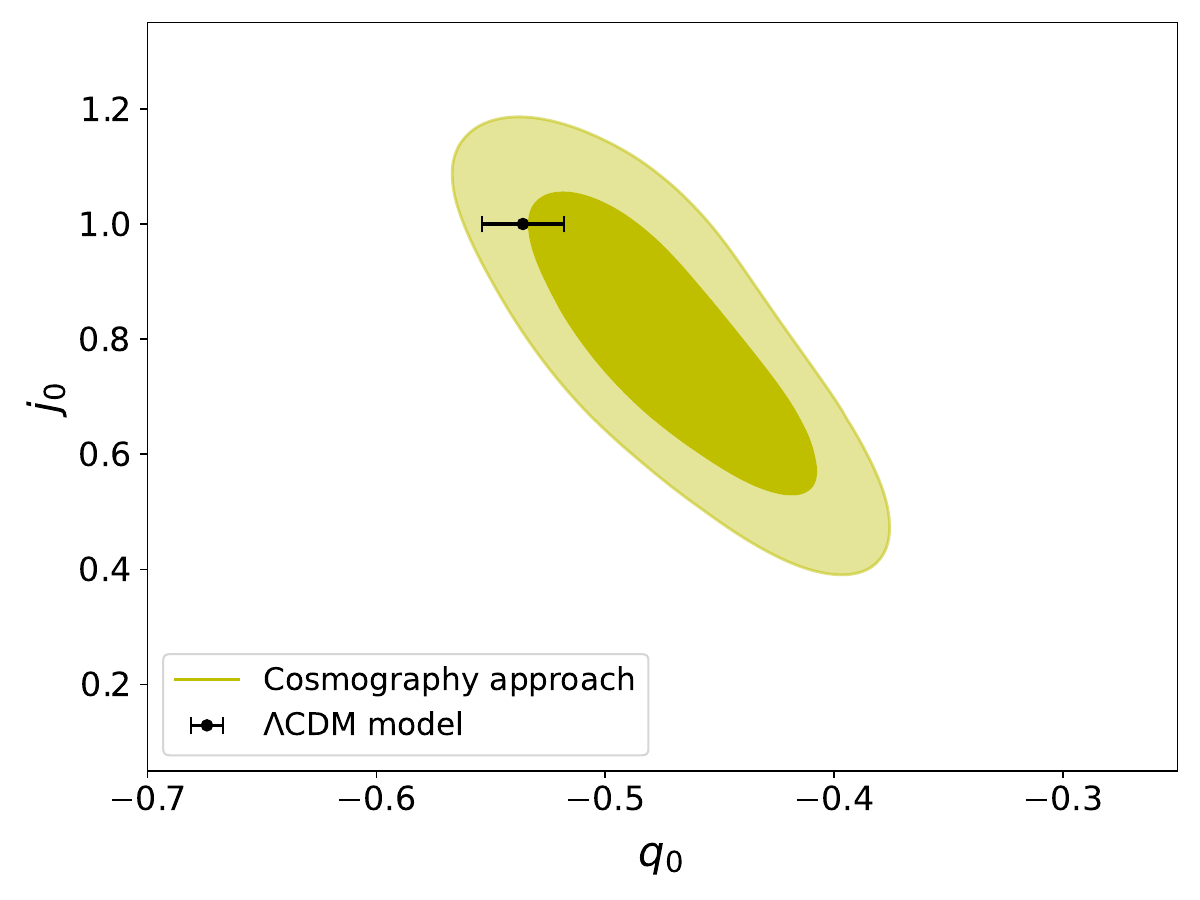}\includegraphics[width=6cm]{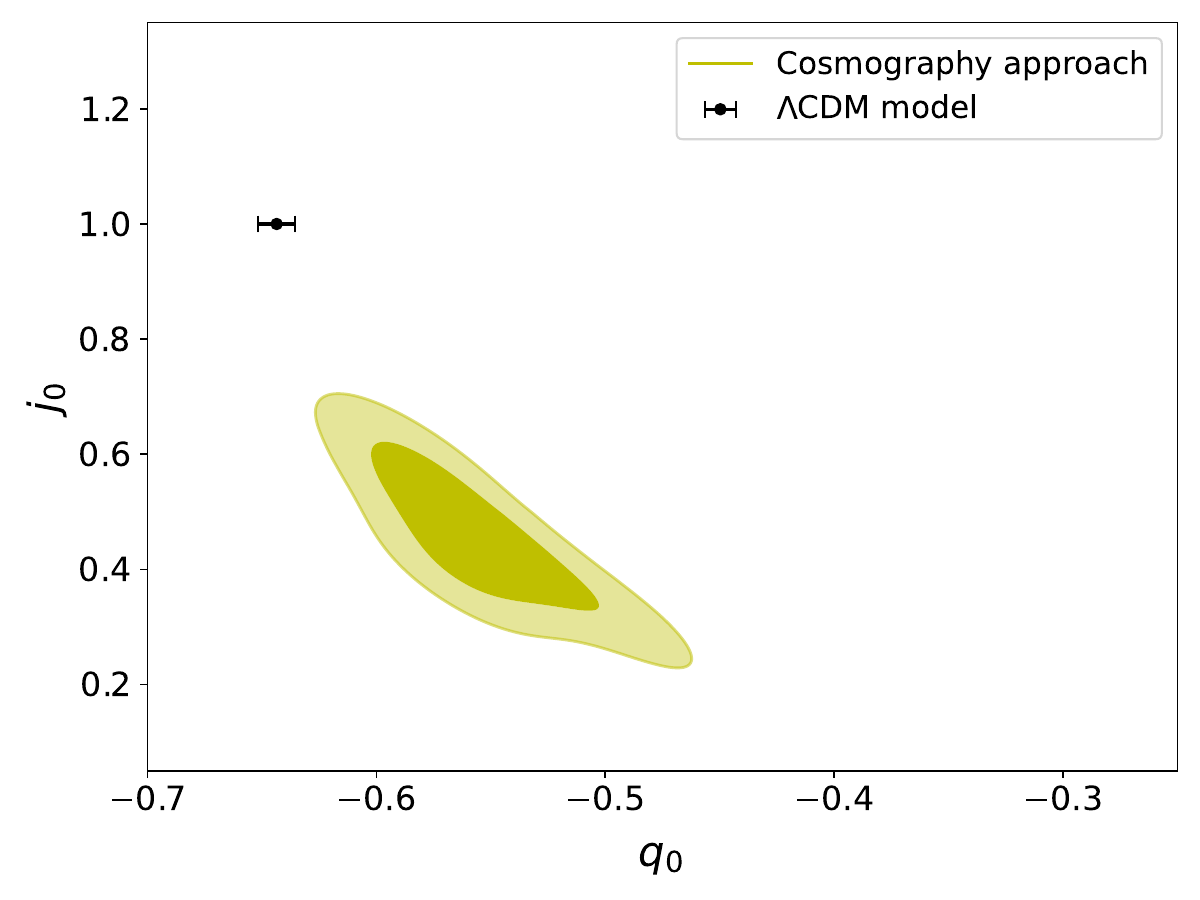}
	\caption{Same as Fig. (\ref{fig2}), but for DESI BAO + Pantheon Plus sample.}
	\label{fig3}
\end{figure*}

\subsection{DESI BAO + Pantheon Plus sample}

In this section, we utilize the combined data of DESI BAO and Pantheon+ SNIa in our analysis. The free parameters in both the flat-$\Lambda$CDM model and the Padé-cosmography are the same as in the previous section. Here, we present our numerical results for the four cases demonstrated earlier. We note that in the Pantheon+ sample, the observed SNIa are distributed at higher redshifts compared to the DES-SN5YR sample.

\textbf{Case 1:} In the first case, we obtain $H_0 = 73.3^{+2.1}_{-3.5}$ km s$^{-1}$ Mpc$^{-1}$ in the flat-$\Lambda$CDM model and $H_0 = 71.2^{+5.5}_{-4.9}$ km s$^{-1}$ Mpc$^{-1}$ in the cosmography approach (see Fig. \ref{fig_H0}). The difference between the $H_0$ values is $0.36\sigma$. Since in this case, we have no Planck prior on $r_d$ and no prior on $M$, the estimated values of $H_0$ have large error bars. In the upper left panel of Fig. \ref{fig3}, we observe that the cosmographic parameters $q_0$ and $j_0$ in the $\Lambda$CDM model are statistically consistent with those obtained using the cosmography approach. Quantitatively, we have $\Delta_{q_0}=1.91\sigma$ and $\Delta_{j_0}=1.41\sigma$ differences, respectively, for the $q_0$ and $j_0$ parameters of the flat-$\Lambda$CDM model and those of the Padé-cosmographic approach. Both deviations are less than $2\sigma$ and can be interpreted as statistical errors. For higher cosmographic parameters, see the first rows of Tables \ref{Tab5} and \ref{Tab6}.

\textbf{Case 2:} In this case, as shown in the second rows of Tables \ref{Tab5} and \ref{Tab6}, and also in Fig. \ref{fig_H0}, the values of $H_0$ in both scenarios are in agreement with the $H_0$ value from the Planck measurement. This result is primarily due to the impact of the Planck prior on $r_d$ while the absolute magnitude $M$ is free. Moreover, the difference between the values of the deceleration parameter $q_0$ and the jerk parameter $j_0$ in the flat-$\Lambda$CDM model and those of the cosmography approach is small enough ($\Delta_{q_0}=0.47\sigma$ for $q_0$ and $\Delta_{j_0}=2.47\sigma$ for $j_0$) to say that the flat-$\Lambda$CDM model and cosmography are in agreement with each other. This result confirms that the Planck prior essentially supports the flat-$\Lambda$CDM model.

\textbf{Case 3:} In this case, since we apply the Cepheid prior on the absolute magnitude $M$ and simultaneously ignore the Planck prior, our constraints on the $H_0$ parameter in both the flat-$\Lambda$CDM model and the cosmographic approach give $H_0 = 73.62 \pm 0.18$ km s$^{-1}$ Mpc$^{-1}$ and $H_0 = 74.22 \pm 0.24$ km s$^{-1}$ Mpc$^{-1}$, respectively. Both values are in complete agreement with the SH0ES result \citep{Riess:2021jrx}. In the cosmographic analysis, we obtain $q_0 = -0.536 \pm 0.018$ in the flat-$\Lambda$CDM model and $q_0 = -0.472^{+0.041}_{-0.046}$ in the Padé-cosmographic approach, indicating a $\Delta_{q_0}=1.36\sigma$ difference between them. Moreover, in the Padé-cosmographic approach, we get $j_0 = 0.79^{+0.21}_{-0.17}$, indicating an $\Delta_{j_0}=1.17\sigma$ deviation from the flat-$\Lambda$CDM value $j_0 = 1$ (see the lower left panel of Fig. \ref{fig3}).

\textbf{Case 4:} Finally, we apply both priors on $r_d$ and $M$ in our analysis. In Fig. \ref{fig_H0} and also the last rows of Tables \ref{Tab5} and \ref{Tab6}, we show that the $H_0$ values for both the flat-$\Lambda$CDM model and the Padé-cosmographic approach are fully consistent with the SH0ES value. This result is similar to what we found in case 4 of the DESI BAO + DES-SN5YR analysis. It is worth noting that the consistency with the SH0ES value is obtained when we use the Planck prior. This means that our analysis is more sensitive to the Cepheid prior when we use both Planck and Cepheid priors simultaneously. In the cosmography analysis, we obtained $q_0 = -0.6436 \pm 0.0081$ and $q_0 = -0.556^{+0.021}_{-0.037}$, respectively, for the flat-$\Lambda$CDM model and the Padé-cosmographic approach, indicating a $\Delta_{q_0}=2.91\sigma$ difference between them. In addition, we obtain $j_0 = 0.456^{+0.082}_{-0.120}$ for the Padé-cosmographic approach, indicating a $\Delta_{j_0}=4.95\sigma$ tension from the constant value $j_0 = 1.0$ in the flat-$\Lambda$CDM cosmology (see the lower right panel of Fig. \ref{fig3}). The tension reported here is roughly similar to the tension reported in case 4 of the DESI BAO + DES-SN5YR analysis. This significant deviation of the standard flat-$\Lambda$CDM model from the cosmographic approach, which is obtained when we apply both Planck and Cepheid priors together, shows that the standard model is an inadequate scenario to handle both aforementioned priors simultaneously.

\begin{figure*} 
	\centering
	\includegraphics[width=8cm]{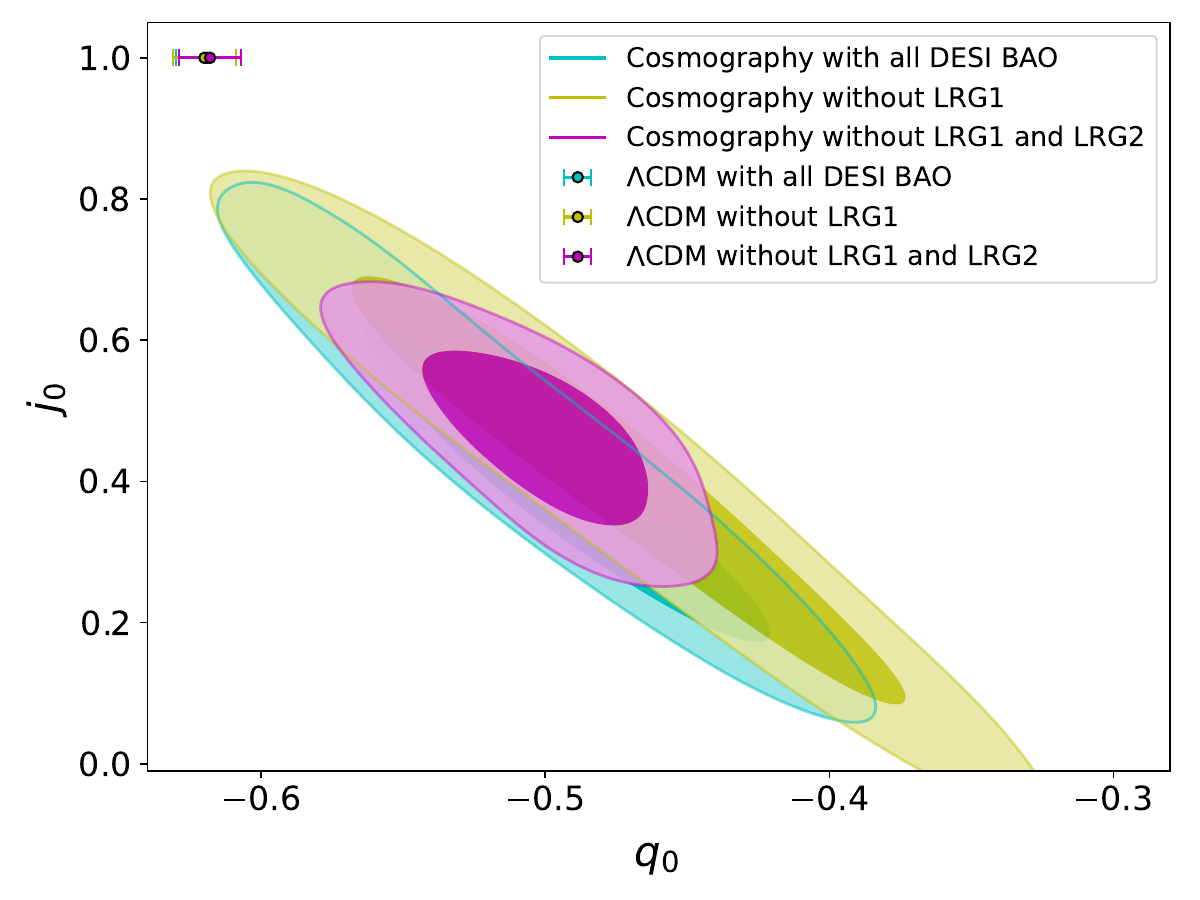}\includegraphics[width=8cm]{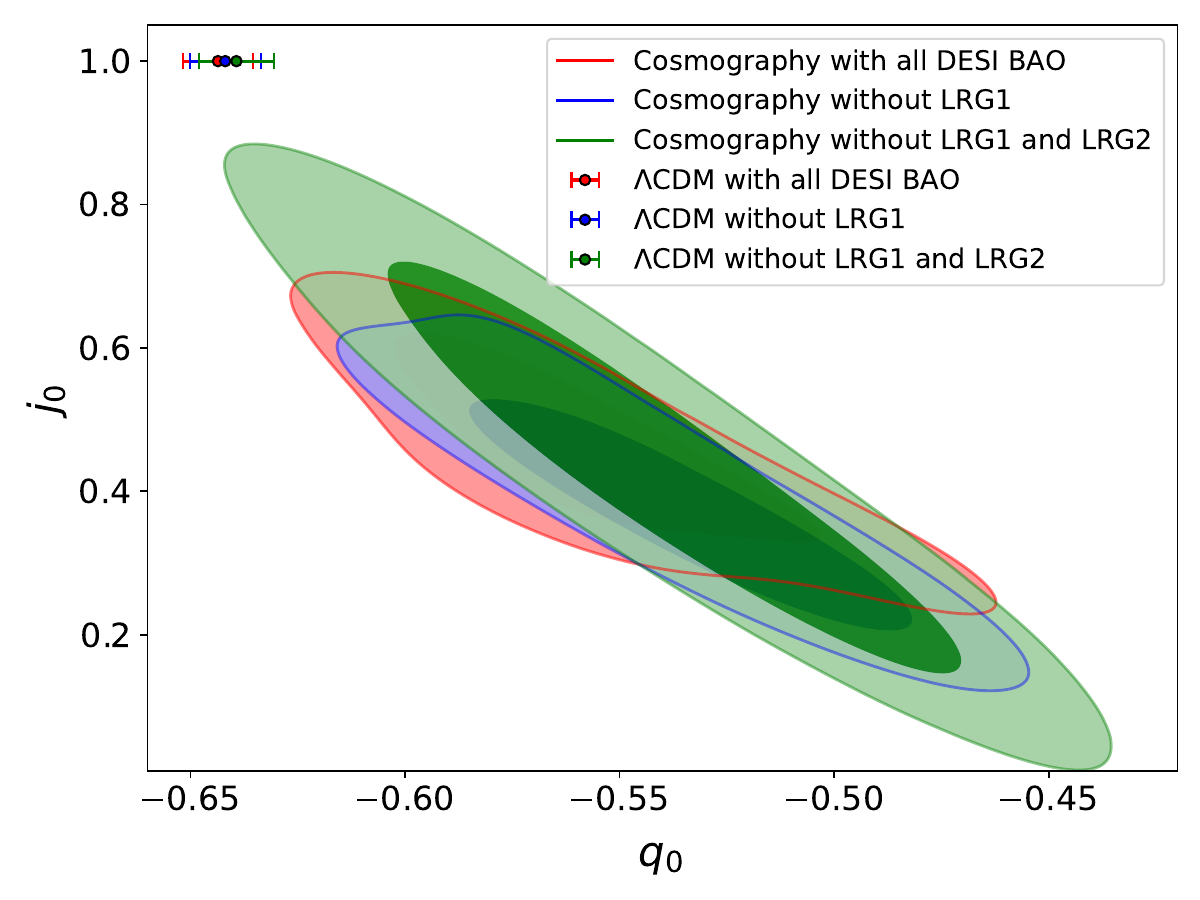}
	\caption{Comparison of cosmographic parameters $q_0$ and $j_0$ constrained in the flat-$\Lambda$CDM model and Padé-cosmography using combined data from DESI BAO + DESY5 (left) and DESI BAO + Pantheon plus sample (right), assuming both Planck and Cepheid priors simultaneously.}
	\label{fig:figlast}
\end{figure*}

\begin{table*}
\centering
\caption{The best-fit values of cosmological parameters with their $1\sigma$ uncertainty, obtained using DESI BAO + DES-SN5YR (upper rows) and DESI BAO + Pantheon plus (lower rows) samples in the flat-$\Lambda$CDM model (left), and the values of cosmographic parameters in the flat-$\Lambda$CDM model calculated using the value of $\Omega_{m0}$ (right).}
\begin{tabular}{c c c | c c c c c}
\hline \hline
Case  & $\Omega_{m0}$ & $H_0$ & $q_0$ & $j_0$ & $s_0$ & $l_0$ & $m_0$\\
\hline
without LRG1 & $0.2536\pm 0.0070$ & $72.97\pm 0.27$ & $-0.620\pm 0.011$ & $1.0$ & $-0.141\pm 0.032$ & $2.630\pm 0.069$ & $-7.97\pm 0.41$\\
\hline
without LRG1 and LRG2 & $0.2548\pm 0.0072$ & $72.95\pm 0.27$ & $-0.618\pm 0.011$ & $1.0$ & $-0.146\pm 0.032$ & $2.641\pm 0.071$ & $-8.04\pm 0.42$\\
\hline
\hline
without LRG1 & $0.2387\pm 0.0055$ & $74.25\pm 0.16$ & $-0.6419\pm 0.0083$ & $1.0$ & $-0.074\pm 0.025$ & $2.486\pm 0.052$ & $-7.12\pm 0.31$\\
\hline
without LRG1 and LRG2 & $0.2405\pm 0.0058$ & $74.25\pm 0.16$ & $-0.6393\pm 0.0087$ & $1.0$ & $-0.082\pm 0.026$ & $2.502\pm 0.055$ & $-7.21\pm 0.32$\\
\hline \hline
\end{tabular}\label{Tab7}
\end{table*}

\begin{table*}
\centering
\caption{The best-fit values of cosmographic parameters with their $1\sigma$ uncertainty, obtained using DESI BAO + DES-SN5YR (upper rows) and DESI BAO + Pantheon plus (lower rows) samples in the Padé-cosmography approach. $\Delta_{q_0}$ and $\Delta_{j_0}$ indicate the deviations of the cosmographic parameters $q_0$ and $j_0$ from those values in $\Lambda$CDM cosmology, respectively.}
\begin{tabular}{c c c c c c c c c}
\hline \hline
& $H_0$ & $q_0$ & $j_0$ & $s_0$ & $l_0$ & $m_0$ & $\Delta_{q_0}$ & $\Delta_{j_0}$\\
\hline
without LRG1 & $72.97\pm 0.55$ & $-0.476^{+0.073}_{-0.065}$ & $0.41^{+0.20}_{-0.25}$ & $-0.170^{+0.073}_{-0.110}$ & $1.61\pm 0.63$ & $-7.4\pm 1.9$  & $2.18 \sigma$ & $2.95 \sigma$\\
\hline
without LRG1 and LRG2 & $73.21\pm 0.36$ & $-0.502^{+0.028}_{-0.021}$ & $0.468\pm 0.084$ & $-0.32\pm 0.10$ & $1.84\pm 0.34$ & $-9.35^{+0.95}{-1.10}$  & $4.32 \sigma$ & $6.34 \sigma$\\
\hline
\hline
without LRG1 & $74.83\pm 0.23$ & $-0.534^{+0.038}_{-0.033}$ & $0.372^{+0.095}_{-0.130}$ & $-0.74^{+0.12}_{-0.10}$ & $0.45^{+0.77}_{-0.87}$ & $-11.9\pm 1.3$  & $2.81 \sigma$ & $5.70 \sigma$\\
\hline
without LRG1 and LRG2 & $74.84\pm 0.26$ & $-0.539^{+0.051}_{-0.043}$ & $0.43^{+0.17}_{-0.23}$ & $-0.55^{+0.45}_{-0.36}$ & $1.47^{+0.51}_{-0.35}$ & $-10.6^{+2.4}_{-3.4}$  & $2.10 \sigma$ & $3.80 \sigma$\\
\hline \hline
\end{tabular}\label{Tab8}
\end{table*}

\subsection{Removing LRG1 and LRG2 data points}

Recent studies on the DESI BAO data sets have shown that the possible deviation from the constant EoS line $w_{\Lambda} = -1$ is mainly due to the LRG1 and LRG2 data points of DESI DR1 BAO measurements (see Table 1 of \citep{DESI:2024mwx}). In the original work, \cite{DESI:2024mwx} updated their analysis by removing DESI BGS and the lowest-redshift LRG1 and LRG2 in $0.6 < z < 0.8$ and replacing them with SDSS data points at $z_{\text{eff}} = 0.15, 0.38$, and $0.51$. They showed that the significance of the tensions with the $\Lambda$CDM model marginally decreases when adopting the (DESI+SDSS) +CMB+SNIa (Pantheon+, Union3, or DES-SN5YR) datasets. A similar result within modified gravity theories was obtained in \citep{Chudaykin:2024gol}, indicating a slightly weaker preference for evolving DE when using the combination of (DESI+SDSS) +CMB+SNIa datasets. The possible deviation from the $\Lambda$CDM model due to LRG1 was also shown within a binned-redshift analysis in \citep{Colgain:2024xqj}. An independent analysis alleviates the deviation from the $\Lambda$CDM model by removing LRG1 and LRG2 data points \citep{Liu:2024gfy}. Additionally, \cite{Ghosh:2024kyd} find inconsistencies between BAO measurements from SDSS and DESI surveys and  both are marginally consistent with the $\Lambda$CDM model in reconstructing the Hubble expansion. While combined datasets supports the $\Lambda$CDM cosmology, indicating a need for further reassessments of DESI survey. Following these efforts, in this section, we remove LRG1 and LRG2 data points and repeat our analysis for case 4, where we join DESI BAO and SNIa from Pantheon+ or DES-SN5YR samples. It is worth mentioning that in case 4, we had both Planck and Cepheid priors simultaneously, and consequently, we observed significant tension between the flat-$\Lambda$CDM model and the cosmographic approach. We first remove LRG1 and then remove both LRG1 and LRG2 from our datasets. The numerical results for the flat-$\Lambda$CDM model are presented in Table \ref{Tab7} and for Padé-cosmography in Table \ref{Tab8}. Additionally, in Fig. \ref{fig:figlast}, we show our results in the $q_0-j_0$ plane for DESI BAO + DES-SN5YR datasets (left panel) and DESI BAO + Pantheon plus (right panel). Numerically, we observe the differences between the flat-$\Lambda$CDM model and the Padé-cosmographic approach as $\Delta_{q_0}=2.18\sigma$ for $q_0$ and $\Delta_{j_0}=2.95\sigma$ for $j_0$ when we utilize the DESI BAO (without LRG1) + DES-SN5YR combination. We obtain larger differences of $\Delta_{q_0}=4.32\sigma$ for $q_0$ and $\Delta_{j_0}=6.34\sigma$ for $j_0$ when utilizing the DESI BAO (without LRG1 and LRG2) + DES-SN5YR combination. In the case of DESI BAO (without LRG1) + Pantheon plus datasets, the differences are $2.81\sigma$ for $q_0$ and $5.70\sigma$ for $j_0$. Finally, in the case of DESI BAO (without LRG1 and LRG2) + Pantheon plus datasets, we obtain $\Delta_{q_0}=2.10\sigma$ for $q_0$ and $\Delta_{j_0}=3.80\sigma$ for $j_0$. We conclude that after removing LRG measurements from the DESI BAO dataset, the tensions between the flat-$\Lambda$CDM model and the Padé-cosmographic approach still exist profoundly. 

\section{Conclusion}\label{conlusion}
In this work, we present a study comparing the standard flat-$\Lambda$CDM model against a cosmographic approach using data from DESI BAO and SNIa compilations, including the DES-SN5YR and Pantheon+ samples. The study aims to test for potential deviations of the standard flat-$\Lambda$CDM model at low redshifts by constraining the cosmographic parameters of the model and comparing them with those of the cosmographic approach. We employed the Padé-cosmographic method \citep{Pourojaghi:2022zrh} as a robust, model-independent approach to reconstruct the Hubble expansion history of the Universe at low redshifts.
When utilizing DESI BAO, DES-SN5YR, or Pantheon+ independently, we observed full consistency between the flat-$\Lambda$CDM model and the cosmographic approach. Additionally, for both the flat-$\Lambda$CDM and cosmographic scenarios, we obtained the Hubble constant ($H_0$) well within the SH0ES range  \citep{Riess:2021jrx}.
 Next, we investigate the impact of Planck prior on the sound horizon at the drag epoch ($r_d$) and Cepheid prior on the absolute magnitude ($M$) of SNIa in our analysis, utilizing combinations of DESI BAO measurements with DES-SN5YR or Pantheon+ samples. We show that when our analysis is free from both priors, we obtain the large errors for our constraints on the cosmological parameters, especially on $H_0$. The large error on $H_0$ value reconciles the Planck-inferred value and SH0ES value of $H_0$. Additionally, in this case, we observe no tension between the flat-$\Lambda$CDM model and cosmographic approach for both combinations of DESI BAO+DES-SN5YR and DESI BAO+Pantheon plus datasets. We then  apply the Planck prior on $r_d$ and consider $M$ to be free. In this case we constrain $H_0$ close to Planck value \citep{Planck:2018vyg}, as expected due to the presence of Planck prior in our analysis, in agreement with the recent work by \cite{DES:2024ywx}. Furthermore, the deviation of the flat-$\Lambda$CDM model from the cosmographic method in $q_0-j_0$ plane is statistical indicating no tension between them.
 In the next step, we relax the prior on $r_d$ and apply the Cepheid prior on $M$. For both data combinations, we obtain the $H_0$ value consistently well within the SH0ES range. This result is due to the presence of Cepheid prior in our analysis without assuming a tight prior on $r_d$. We also observe consistency between the flat-$\Lambda$CDM model and cosmographic approach in $q_0-j_0$ plane when utilizing the DESI BAO+DES-SN5YR combination. However, for DESI BAO+Pantheon plus combination, we observe considerable tension between the model and the cosmographic approach. This may be due to extension of the SNIa data points in Pantheon+ sample to higher redshifts compared to DES-SN5YR sample. Finally, we apply both Planck and Cepheid priors simultaneously in our analysis. For both data combinations, we measure $H_0$ consistently well with the SH0ES value. However, we observe significant tension between the flat-$\Lambda$CDM model and  the cosmographic approach in $q_0-j_0$ plane and in higher cosmographic parameters for both DESI BAO+DES-SN5YR and DESI BAO+Pantheon plus data combinations. This inconsistency suggests that the standard flat-$\Lambda$CDM model is not consistent with both high-redshift Planck CMB observations and local measurements of Cepheids at the same time. We repeat our analysis for the last case where we have both priors, by removing LRG1 and LRG2 from DESI BAO measurements. Interestingly, for both data combinations assumed in our analysis, the tension between flat-$\Lambda$CDM and cosmographic approach is more pronounced. This result suggests the potential for new physics or the need for modifications to the standard $\Lambda$CDM model. A straightforward modification could involve extending the $\Lambda$CDM model by introducing dynamical dark energy (DE) scenarios with an evolving equation of state (EoS) parameter, $w_{de} \neq -1$. Additionally, any deviation from the standard model can be interpreted as a general modification of General Relativity (GR) through the window of modified gravity theories. Alternatively, this deviation might be due to unrecognized systematic errors in the observational data or analysis methods, which should be addressed in future investigations.

\textbf{Acknowledgment}\\
We are grateful to the anonymous referee for their thorough review of our manuscript and for providing constructive and valuable comments that significantly improved our paper. The work of SP is based upon research funded by the Iran National Science Foundation (INSF) under project No. 4024802. ZD was supported by  the Korea Institute for Advanced Study Grant No 6G097301.\\

\textbf{Data Availability}\\
The data used in this work can be obtained upon reasonable
request.
\newpage

\bibliographystyle{mnras}
\bibliography{ref}
\label{lastpage}

\appendix
\section{Robustness of Padé(3,2) up to $z = 2.5$}
\label{app:apx1}
\begin{figure*} 
	\centering
	\includegraphics[width=5.95cm]{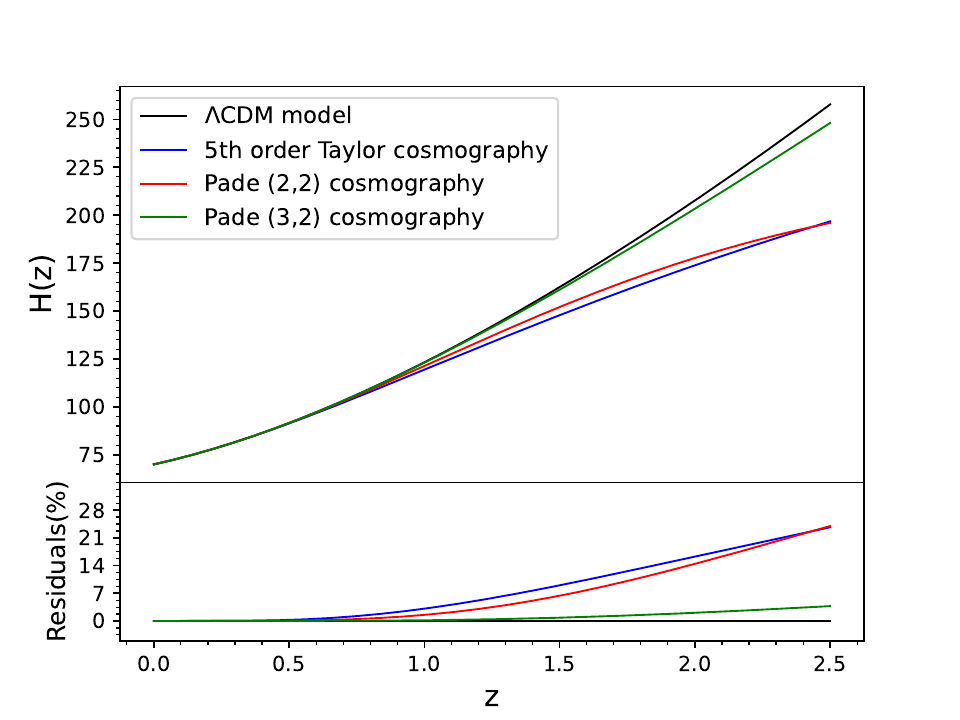}
        \includegraphics[width=5.95cm]{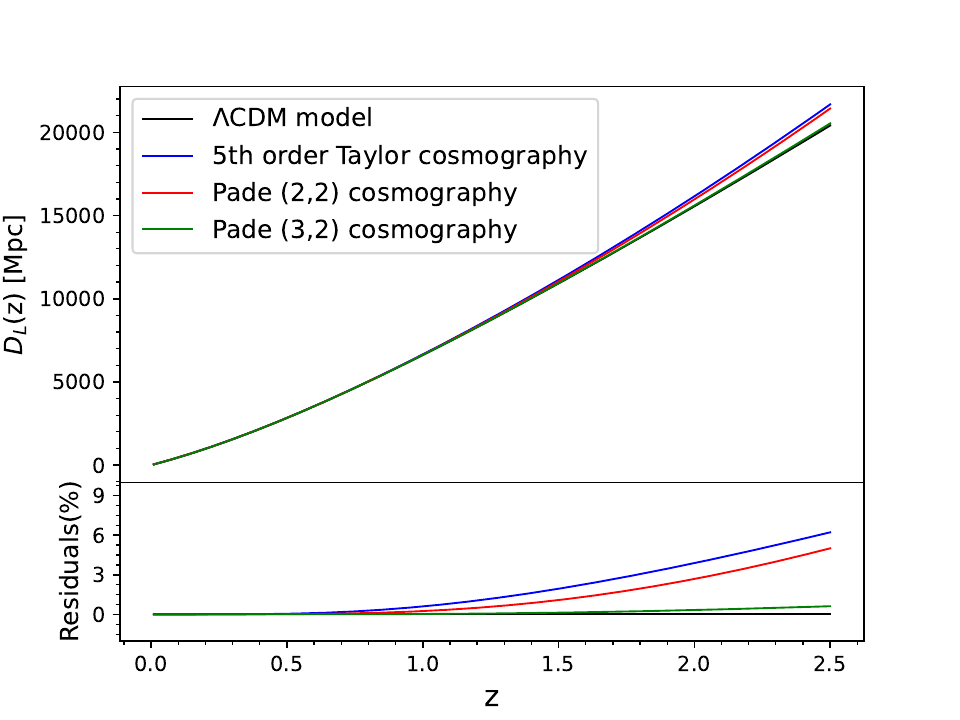}
        \includegraphics[width=5.95cm]{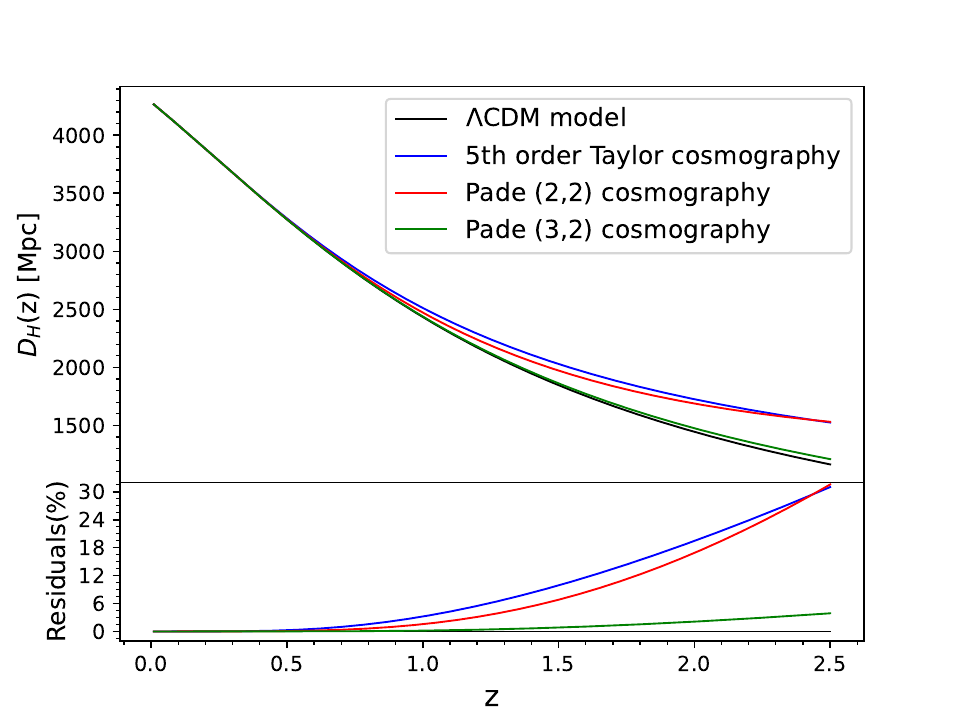}
        \includegraphics[width=5.95cm]{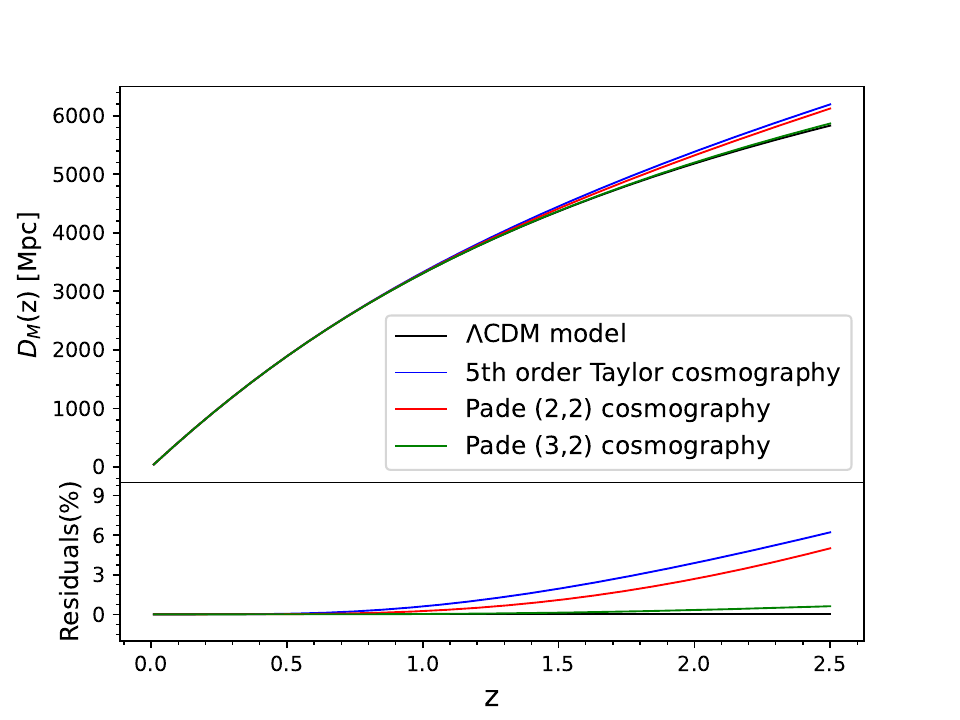}
	\caption{The redshift evolution of the reconstructed Hubble parameter (upper-left), luminosity distance (upper-right), Hubble distance (lower-right) and comoving distance (lower right) for different Padé(2,2), Padé(3,2) and $5th$-order Taylor approximations and deviations from those of the standard $\Lambda$CDM cosmology.}
	\label{fig:fig1}
\end{figure*}

For a typical value of $\Omega_{m0}=0.3$, we first directly calculate the Hubble parameter via Eq.~(\ref{eq:c6}), and cosmographic parameters via Eq.~(\ref{c7}) for the standard flat-$\Lambda$CDM cosmology. Substituting the cosmographic values into Eqs.~(\ref{eq5}), (\ref{eq:eq3}), and (\ref{eq:eq4}), we reconstruct the Hubble parameter of the standard $\Lambda$CDM cosmology using the Padé(3,2) approximation. We then reconstruct the comoving distance, $D_M$, Hubble distance, $D_H$, and luminosity distance, $D_L$, by substituting Eq.~(\ref{eq:eq3}) into Eqs.~(\ref{eq:dM}), (\ref{eq:dH}), and (\ref{eq:P4}), respectively. In these substitutions, we use the cosmographic values of the standard $\Lambda$CDM model with $\Omega_{m0}=0.3$ and $H_0=70 \, \text{km/s/Mpc}$.
Additionally, we perform the same exercise for the Padé(2,2) and Taylor series (up to the 5th order). Note that in Padé(2,2), we have one less free parameter since $P_3=0$, while the number of free parameters in the 5th order Taylor series is equal to that in Padé(3,2). For more details, we refer readers to \citep{Pourojaghi:2022zrh}. In Fig.~\ref{fig:fig1}, we show the evolution of the Hubble parameter $H(z)$ (upper-left panel), luminosity distance $D_L(z)$ (upper-right panel), Hubble distance $D_H(z)$ (lower-left panel), and comoving distance $D_M(z)$ (lower-right panel) in terms of redshift for different Padé and Taylor approximations, comparing them with those of the standard flat-$\Lambda$CDM cosmology. 
At redshifts $z \leq 1$, all approximations can effectively reconstruct the Hubble parameter and various distances with small and negligible truncation errors. However, at higher redshifts, we observe that the Padé(2,2) and 5th-order Taylor series cannot accurately reconstruct the Hubble parameter or the various comoving, Hubble, and luminosity distances of the $\Lambda$CDM model. The large deviations between the reconstructed quantities of Padé(2,2) and the 5th-order Taylor series compared to those of the $\Lambda$CDM model arise from significant truncation errors in these approximations. Conversely, Padé(3,2) performs well up to redshift $z \simeq 2.5$, with a maximum truncation error of around 4\% for the reconstructed Hubble and Hubble distance parameters, and 0.6\% for the reconstructed luminosity and comoving distances. The numerical values for the percentage differences between the reconstructed quantities in different cosmographic approaches and those of the standard $\Lambda$CDM cosmology at redshifts $z=1.0$ and $z=2.5$ are presented in Table \ref{Tab:per}. Therefore, it is worthwhile to acknowledge the robustness of Padé(3,2) and apply it in the cosmographic approach for redshifts smaller than $z \simeq 2.5$, where we have DESI BAO and SNIa observations.

\begin{table}
    \centering
    \caption{Percentage difference between reconstructed quantities in different cosmographic approaches and those of the standard $\Lambda$CDM cosmology.}
        \begin{tabular}{c c c c c c}
        \hline \hline
        $approximation$ & $redshift$ & $H$ & $D_H$ & $D_M$ & $D_L$\\
        \hline
        \multirow{2}{*}{$5th$-order Taylor} & $z=1$ & $3.1\%$ & $3.2\%$ & $0.6\%$ & $0.6\%$ \\
        & $z=2.5$ & $23.7\%$ & $31.0\%$ & $6.2\%$ & $6.2\%$ \\
        \hline
        \multirow{2}{*}{Padé(2,2)} & $z=1$ & $1.5\%$ & $1.6\%$ & $0.2\%$ & $0.2\%$ \\
        & $z=2.5$ & $24.0\%$ & $31.5\%$ & $5.0\%$ & $5.0\%$ \\
        \hline
        \multirow{2}{*}{Padé(3,2)} & $z=1$ & $0.2\%$ & $0.2\%$ & $0.03\%$ & $0.03\%$ \\
        & $z=2.5$ & $3.8\%$ & $3.9\%$ & $0.6\%$ & $0.6\%$ \\
        \hline \hline    
    \end{tabular}\label{Tab:per}
\end{table}
\end{document}